%% file: Enhanced_gravitational_entanglement_via_modulated_optomechanics.tex
\documentclass[twocolumn,amsmath,amssymb,superscriptaddress,a4paper,accepted=2023-10-15]{quantumarticle}

\pdfoutput=1
\usepackage[numbers]{natbib}

\usepackage{dcolumn}
\usepackage{bm}
\usepackage{float}
\usepackage[caption = false]{subfig}
\usepackage{graphics,xcolor,import}
\usepackage{braket}
\usepackage{comment}
\usepackage{mathtools}
\usepackage{dsfont}
\usepackage{wrapfig}
\usepackage{microtype}
\usepackage{multirow}
\usepackage{booktabs}
\usepackage{hyperref}
\usepackage{subdepth}
\usepackage[normalem]{ulem}
\usepackage{cellspace}
\usepackage{tabularx,ragged2e,booktabs}
\usepackage[bottom]{footmisc}

\newcommand{\Tr}{\mathrm{Tr}}
\newcommand{\tr}{\mathrm{tr}}
\newcommand{\ad}{\hat{a}^{\dag}}
\newcommand{\bd}{\hat{b}^{\dag}}
\renewcommand{\b}{\hat{b}}
\newcommand{\avg}[1]{\langle #1\rangle}
\newcommand{\ketbra}[2]{\ket{#1}\!\!\bra{#2}}

\usepackage{graphicx}
\newcount\colveccount
\newcommand*\colvec[1]{
	\global\colveccount#1
	\begin{pmatrix}
		\colvecnext
	}
	\def\colvecnext#1{
		#1
		\global\advance\colveccount-1
		\ifnum\colveccount>0
		\\
		\expandafter\colvecnext
		\else
	\end{pmatrix}
	\fi
}

\begin{document}

	\title{Enhanced Gravitational Entanglement via Modulated Optomechanics}
	
	\author{A. Douglas K. Plato}
	\email{adouglaskplato@gmail.com}
	\affiliation{Institut f{\"u}r Physik, Universit{\"a}t Rostock, Albert-Einstein-Stra{\ss}e 23, 18059 Rostock, Germany}

	\author{Dennis R\"atzel}
    \affiliation{ZARM, University of Bremen, Am Fallturm 2, 28359 Bremen, Germany}
	\affiliation{Institut f\"ur Physik, Humboldt Universit\"at zu Berlin, Newtonstraße 15, 12489 Berlin, Germany}
	
	\author{Chuanqi Wan}

\begin{abstract}
		The role of entanglement in determining the non-classicality of a given interaction has gained significant traction over the last few years. In particular, as the basis for new experimental proposals to test the quantum nature of the gravitational field. Here we show that the rate of gravity mediated entanglement between two otherwise isolated optomechanical systems can be significantly increased by modulating the optomechanical coupling. This is most pronounced for low mass, high frequency systems -- convenient for reaching the quantum regime -- and can lead to improvements of several orders of magnitude, as well as a broadening of the measurement window. Nevertheless, significant obstacles still remain. In particular, we find that modulations increase decoherence effects at the same rate as the entanglement improvements. This adds to the growing evidence that the constraint on noise (acting on the position d.o.f) depends only on the particle mass, separation, and temperature of the environment and cannot be improved by novel quantum control. Finally, we highlight the close connection between the observation of quantum correlations and the limits of measurement precision derived via the Cram\'er-Rao Bound. An immediate consequence is that probing superpositions of the gravitational field places similar demands on detector sensitivity as entanglement verification.
\end{abstract}

\maketitle

\twocolumngrid
%---------------------------------
\section{Introduction}\label{Sec:intro}
One of the central features in most attempts at combining gravity and quantum theory is the assumption that the gravitational field should be quantised. However, in contrast to the other known forces, there has been no experimental evidence to motivate this approach. Instead the reliance is based largely on a combination of aesthetic appeal along with a variety of well known inconsistency arguments for coupling classical and quantum fields \cite{Eppley1977,Peres2001,Terno2006}. As such, the question of experimental verification, and indeed whether this is in principle even possible, is of significant importance. 

In recent years one common approach is to focus on identifying ways that phenomenological features, such as a minimal length or modified dispersion relation, may be accessible in the low energy regime. Surprisingly, this has led to a number of proposals for so-called table-top experiments, in particular within the framework of atomic, molecular and optical (AMO) physics \cite{Amelino-Camelia2009,Bekenstein2012,Pikovski2012,Anastopoulos2015,Carlesso2019,Albrecht2014}, (see also \cite{Plato2016} for a recent review). In these approaches, one attempts to effectively reach the quantum gravity regime through the use of composite systems, provided that large mass quantum states can be produced. 

An alternative strategy is to bypass theory specific predictions and instead tackle the question of nonclassicality directly. Recently it has been argued that witnessing the build up of quantum correlations through only a gravitational interaction is sufficient to rule out a purely classical description of gravity \cite{Bose2017:spin,Marletto2017,Wan2017}. This observation rests on the fact that Local Operations and Classical Communication (LOCC) cannot increase entanglement between two systems \cite{Chitambar2014}. From this perspective, entanglement verification can be viewed as a higher level test of the underlying theory.

One can expect that the detection of quantum features should depend heavily on the sensitivity of the test system to gravity. This requires not only that the field is measurable, but that the precision should be high enough to detect quantum fluctuations of the source. It is well known that by employing resonant driving, measurement sensitivity can be significantly improved. A promising setup is then provided by two disconnected optomechanical systems placed in close proximity. Here, radiation pressure induces oscillations in the mechanical elements, which if sufficiently isolated will only be coupled through their gravitational interaction. Individually, such systems serve as promising candidates for gravity sensors \cite{Matsumoto2019,Catano-Lopez2020,RademacherMillenLi+2020+227+239,montoya2022scanning}, and the fundamental limits on their precision in the nonlinear regime have previously been explored \cite{Armata2017, Qvarfort2018}. An important observation is that a much greater sensitivity can be achieved if, in addition, the optomechanical coupling is modulated at resonance with both the mechanical and driving frequency \cite{Qvarfort2021}. 

This motivates an extension to an optomechanical experiment first suggested by two of us to entangle the motion of two movable mirrors through their mutual gravitational field \cite{Wan2017}. We show that if gravity can be modeled by a quantised Newtonian interaction then modulations of the optomechanical coupling close to the mechanical resonance can significantly enhance the rate of entanglement generation. Such modulations have been experimentally demonstrated, for example, in nanomechanical setups~\cite{szorkovszky_mechanical_2011} or in levitated optomechanical systems by employing hybrid-Paul traps~\cite{Millen:2015cav,Fonseca:2016non,aranas_split-sideband_2016}.
In particular, we find that entanglement can grow asymptotically with the cube of the interaction time, and proportional to the variance of the photon number operator of the initial cavity state (corresponding to roughly the position variance of the mirrors). 

In fact, the link between measurement precision and the generation of quantum correlations can be made more concrete and we will show that the integration time needed for an optomechanical sensor to achieve a measurement precision to better than the quantum variance of the corresponding operator (in our case the mirror position or cavity photon number) is closely related to the entanglement time. From this perspective, the development of (decoherence free) quantum systems as sensors is as important as creating large nonclassical states. In particular, this implies that proposals to test the superposition principle for gravitational fields \cite{Anastopoulos2015, Carlesso2019} are operationally just as difficult as entanglement verification. It also emphasises that one does not always need to resort to strict gravitational cat states in order to test for nonclassical features of gravity. 

In this paper we analyse the feasibility of performing such an experiment in a regime where entanglement would be expected. We start by introducing a model Hamiltonian which describes the low energy gravitational interaction of two optomechanical systems in the nonlinear regime (sections \ref{sec:NewtonianInteraction} and \ref{Sec:Model}). Solving the dynamics for arbitrary time dependent parameters leads to a coupling between the cavity fields proportional only to the respective photon number operators. This allows for a simple estimate of the entanglement rate, provided measurements can made when the combined field-field state is approximately pure (section \ref{Sec:EntanglementTime}). 

We then quantify the effects of extraneous couplings through the use of a suitable entanglement witness (section \ref{sec:verification}). This reveals a trade off between the width of the viable measurement window and the initial cooling (previously noted for the interference visibility \cite{Marshall2003}), but also exposes additional demands on frequency matching of the two mechanical elements. For high photon numbers, these constraints can be significant, though we show that at least for the first, the temperature dependence can be reduced by using purely oscillatory modulations.

On the other hand, decoherence requirements are severe and cannot be mitigated through local control of the dynamics. We recover known noise conditions on the mechanical relaxation rate in the limit of perfect measurement sensitivity \cite{Kafri_2014,miao2020quantum}, as well as the more stringent bounds needed to keep the witness within some fraction of its optimal value. In both cases these depend only on the mass, equilibrium separation and the temperature of the environment and suggest improvements of many orders of magnitude are needed before such an experiment can be performed.

Further experimental limitations due to Casimir effects, stray charges and restrictions on the mechanical motion are then discussed in sections \ref{Sec:Casimir} and \ref{Sec:Limitations}. Finally, we contrast our results with those obtained through a semi-classical analysis and highlight the connections to quantum metrology in sections \ref{Sec:semiclassical} and \ref{Sec:Metrology}.

\section{Low-energy gravitational interaction}\label{sec:NewtonianInteraction}
For lab based experiments, where the typical energy scale is small, a common approach is to start by quantising the classical Newtonian potential \cite{Anastopoulos_2014,Kafri_2014,Wan2017, Anastopoulos_2020}, $V_G=-\frac{Gm_1m_2}{r}$, where $G$ is the gravitational constant and $m_i$ are the masses of two particles separated by distance $r$. Intuitively this can be done by introducing the quantised particle positions $\hat{x}_1$ and $\hat{x}_2$ as small perturbations around an equilibrium distance $d$, such that $r \rightarrow \hat{r}=d+\hat{x}_2-\hat{x}_1$. However a similar result can also be obtained from the gravitational interaction of two scalar fields in perturbative quantum gravity when restricted to the two-particle sector (in the non-relativistic limit)\footnote{The derivation becomes non-trivial if the composite nature of the interacting systems has to be taken into account. For example, a field theoretical treatment of a gravitationally self-interacting BEC in a double-well potential has been presented in \cite{Anastopoulos_2020}. Here, we simply assume that the strong interaction of each mechanical element's components (atoms, free electron gas etc.) lead to the gravitational coupling being effectively via their center-center-of-mass degrees of freedom in the form of $\hat{H}_G$.} \cite{Anastopoulos_2014}. In both cases this leads to a Hamiltonian term,
\begin{equation}\label{eq:gravitationalpotentialfull}
\hat{H}_G=- \frac{Gm_1m_2}{|d+\hat{x}_2-\hat{x}_1|} .
\end{equation}

Provided the expected displacements are small, we can expand to second order in $\hat{x}_2-\hat{x}_1$,
\begin{equation}\label{eq:gravitationalpotential}
\hat{H}_G \approx - \frac{Gm_1m_2}{d} \left(1 - \frac{\hat{x}_2-\hat{x}_1}{d} + \frac{(\hat{x}_2-\hat{x}_1)^2}{d^2} \right).
\end{equation} 
The last term in particular leads to a quantised gravitational interaction $\hat{H}_{int}=\frac{2Gm_1m_2}{d^3}\hat{x}_1\hat{x}_2$, and enables the generation of entanglement between the two particles. In general this interaction will not commute with the local dynamics, which in turn leads to non-trivial evolution on the relevant timescale for entanglement generation. The simplest example is given by two free particles with $\hat{H}_{0,i}=\hat{p}^2_i/2m_i$, where the additional second order terms in (\ref{eq:gravitationalpotential}) give rise to local Hamiltonians for (shifted) inverted harmonic oscillators,
\begin{equation}\label{2particleHamiltonian_Newtonian}
\hat{H}=\frac{\hat{p}_1^2}{2m_1} + \frac{\hat{p}_2^2}{2m_2} - \gamma \hat{x}_1^2 -\gamma \hat{x}_2^2 + 2\gamma \hat{x}_1\hat{x}_2 +\gamma d (\hat{x}_2-\hat{x}_1)
\end{equation}
where $\gamma=Gm_1m_2/d^3$. 

Here the behaviour of entanglement generation can be markedly different if the sign of the potential is reversed \cite{Sudhir2012,Krisnanda2020}, and in particular tends to be lower (or indeed can return to zero) when the particles are trapped in harmonic potentials. In practice, however, it can be beneficial to maintain localisation through the introduction of an additional potential (which can be either time dependent \cite{Bose2017:spin} or independent \cite{Wan2017}). One can then attempt to offset the reduced entanglement rate by transferring the correlations to a set of ancilla states in such a way that they increase after every cycle. The advantage of this approach is that entanglement can often be more easily verified between the ancilla degrees of freedom, either because they are chosen to have a reduced dimension or because they are easier to access in an experiment. This type of setup forms the basis for the majority of proposals to test gravitationally generated entanglement.

%---------------------------------
\section{Model}\label{Sec:Model}
A practical realisation of the above can be provided by optomechanical systems, in which the entanglement is transferred to optical degrees of freedom via a dispersive coupling of the form $g(t)\hat{N} \hat{x}$ (where $\hat{N}$ is the photon number operator). We will now show that for two coupled cavities, the gravity mediated photon-photon coupling takes on a surprisingly simple form: $\hat{U}_{12}(t)=e^{iD(t) \hat{N}_1\hat{N}_2}$. The reduced complexity of the interaction simplifies the analysis of the entanglement rate, and motivates an optimal choice for the initial cavity state. In practice however, this is not easily achieved. Instead we argue that by tailoring the time dependence of the optomechanical coupling $g(t)$, one can significantly improve the performance of less optimal (but easier to produce) cavity states. 
\begin{figure}[htb]
	\def\svgwidth{\columnwidth}
	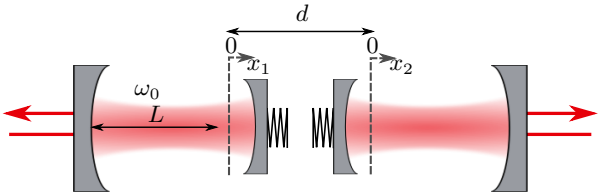
	\caption{\label{fig:model1} Two Fabry-P\'erot cavities brought in close proximity. The moving elements interact via their gravitational fields. To second order this leads to an $\hat x_1 \hat x_2$ coupling and allows the generation entanglement between the mechanical modes. This is exchanged back and forth with the cavity fields, which themselves become entangled around multiples of the mechanical period $1/\omega_m$.   }
\end{figure}

We begin by assuming two optomechanical cavities coupled only through the gravitational interaction of their respective mechanical elements. This can be realised, for example, in a Whispering Mode Gallery system where the mechanical elements are made up of fuse-silica microspheres. An alternate setup is shown in Fig \ref{fig:model1}, using Fabry-P\'erot cavities, but in principle the analysis can be readily applied to a range of systems. In general, the precise form of the interaction will depend on the geometry of the masses, $m_1$ and $m_2$, however for simplicity we assume these to be by point sources (\ref{eq:gravitationalpotential}). If each cavity contains only a single mode, then the total Hamiltonian describing this system is given by,
\begin{equation}\label{eq:systemHamiltonian}
\begin{aligned}
\hat{H}&=\sum_{j=1}^2 \hbar \omega_{0,j} \hat{N}_j +\frac{\hat{p}_j^2}{2m_j} + \frac{1}{2}m_j \tilde{\omega}_{j}^2\hat{x}_j^2 \\
& \quad +\sum_{j=1}^2(-1)^j(g_j(t) \hat{N}_j+S_j(t))\hat{x}_j +2\gamma\hat{x}_1\hat{x}_2 \,,
\end{aligned}
\end{equation}
where $\hat{N}_j=\ad_j\hat{a}_j$ are the number operators for the respective cavity fields (with frequency $\omega_{0,j}$), $\tilde{\omega}_j^2 = \omega_{m,j}^2-\frac{2\gamma}{m_j}$ are the mechanical frequencies shifted by the inverted harmonic potential coming from (\ref{eq:gravitationalpotential}), and $S_j(t)$ are arbitrary linear displacements. The latter contain the first order gravitational contribution in (\ref{eq:gravitationalpotential}), but could also include effects from additional
external potentials. Note that the relative minus sign between cavities accounts for the opposite alignment. 

The time evolution corresponding to (\ref{eq:systemHamiltonian}) can most easily be solved by first decoupling the mechanical degrees of freedom. This can be achieved via the symplectic transformations \cite{Sudhir2012},
\begin{equation}\label{symplectictransforms}
\begin{aligned}
\hat{x}_1 &= c\hat{x}_+ - b s\hat{x}_-;\qquad \hat{p}_1 = c\hat{p}_+ - \frac{1}{b} s\hat{p}_-,\\
\hat{x}_2 &= c\hat{x}_- + \frac{1}{b}s\hat{x}_+;\qquad \hat{p}_2 = c\hat{p}_- + bs\hat{p}_+,
\end{aligned}
\end{equation}
where $c\equiv\cos(a)$ and $s\equiv\sin(a)$, provided that the parameters $a$ and $b$ satisfy,
\begin{equation}\label{diagonalisationcondition}
\tan{2a}=\frac{   4 \gamma }{\sqrt{m_1m_2}(\tilde{\omega}_1^2-\tilde{\omega}_2^2)}, \qquad b=\sqrt{\frac{m_2}{m_1}}.
\end{equation}
It is convenient to rewrite the transformed mechanical degrees of freedom in terms of the mode operators, $\hat{b}_\pm$ and $\bd_\pm$ via $\hat{x}_\pm=x_{0,\pm}  (\bd_\pm + \hat{b}_\pm)$ (and similarly for $\hat{p}_\pm$), 
where $x_{0,+}=\sqrt{\frac{\hbar}{2m_1 \omega_+}} $,  $x_{0,-}=\sqrt{\frac{\hbar}{2m_2\omega_-}}$, and,
\begin{equation}\label{frequencies}
\omega^2_\pm=\frac{1}{2}\left(\tilde{\omega}_1^2 + \tilde{\omega}_2^2 
\pm\sqrt{\frac{16\gamma^2}{m_1m_2} + (\tilde{\omega}_1^2-\tilde{\omega}_2^2)^2} \right).
\end{equation}
We note for later, that when $\omega_{m,1}=\omega_{m,2} \equiv \omega_m$ the frequencies simplify to $\omega^2_+ = \omega_m^2$ and $\omega^2_- = \omega_m^2(1-4\gamma_g)$ where $\gamma_g=\frac{1}{2}\frac{\gamma(m_1+m_2)}{ m_1m_2 \omega_m^2}$. So far the parameter dependence in each cavity is arbitrary, however we now impose the restriction $g_j(t)=g(t)\bar{g}_j$. This allows us to define time independent combinations of the number operators associated to each plus/minus mode. The Hamiltonian (\ref{eq:systemHamiltonian}) can then be written as $\hat{H}=\sum_{j=1}^2\hbar \omega_{0,j}\hat{N}_j+ \hat{H}_+ +\hat{H}_-$ where,
\begin{equation}\label{eq:modulatedH}
\begin{aligned}
	\hat{H}_\mu(t)&= \hbar \omega_\mu \hat{b}_\mu^{\dagger} \hat{b}_\mu  \\
	&  \quad - \hbar \omega_\mu \left(k_\mu(t) \hat N_\mu + \mathcal{D}_{1,\mu}(t)\right)\left(\hat{b}_\mu^\dagger + \hat{b}_\mu\right), 
\end{aligned}
\end{equation}
(for index $\mu=+,-$), with the transformed number operators,
\begin{equation}
\begin{aligned}
\hat{N}_+&=\sqrt{2}(\bar{g}_1 c \hat{N}_1 - \bar{g}_2 \frac{s}{b}\hat{N}_2),\\
\hat{N}_-&=-\sqrt{2}(bs\bar{g}_1\hat{N}_1 + c\bar{g}_2 \hat{N}_2),
\end{aligned}
\end{equation}
and $k_\pm(t)=x_{0,\pm}g(t)/(\sqrt{2}\hbar \omega_\pm)$. Similarly, the (potentially) time dependent displacement terms are defined by $\mathcal{D}_{1,+}(t) = \frac{x_{0,+}}{\hbar \omega_+}(cS_1-\frac{s}{b}S_2)$ and $\mathcal{D}_{1,-}(t) = -\frac{x_{0,-}}{\hbar \omega_-}(bs S_1 + c S_2)$. In general these terms do not contribute to the entanglement, though they can play a role in determining the allowable parameter regime (see section \ref{Sec:Limitations}).

We now observe that, 
\begin{equation}
[\hat{H}_\mu(t),\hat{H}_{\mu'}(t')]=0=[\hat{N}_j,\hat{H}_\mu], 
\end{equation}
for $j=1,2; \,\mu,\mu'=+,-$; $\,\mu\neq\mu'$, and so the the evolution operator corresponding to (\ref{eq:systemHamiltonian}) can be split into,
\begin{equation}
\hat U(t)=e^{-i(\omega_{0,1}\hat N_1+ \omega_{0,2}\hat N_2)t}\hat U_{+}(t)\hat U_{-}(t),
\end{equation}
where each $\hat U_\mu(t)$ corresponds to the individual evolution of the Hamiltonians (\ref{eq:modulatedH}). A general ansatz can be found by considering the closed Lie-Algebra generated by the terms in (\ref{eq:modulatedH}) \cite{qvarfort2020time},
\begin{equation}\label{eq:Uk}
\hat{U}_\mu =  e^{-i\omega_\mu t \,\hat{b}^\dagger_\mu \hat{b}_\mu}  e^{i\theta_\mu}e^{i(A_\mu\hat{N}_\mu +B_\mu\hat{N}_\mu^2)}\hat{D}_\mu (C_\mu+K_{\hat{N}_\mu}\hat{N}_\mu) ,
\end{equation}
where $\hat{D}_\mu(\alpha)=\exp(\alpha \bd_\mu - \alpha^* \hat{b}_\mu)$ is the displacement operator acting on the transformed mechanical modes, and the time dependence of the coefficients have been suppressed for notational convenience. Transforming back to the physical cavity field modes, we find the total evolution operator takes the form,
\begin{equation}
\begin{aligned}\label{Ugeneral}
\hat{U}(t)&=e^{i\theta}   e^{-i(\omega_{0,1}\hat N_1 + \omega_{0,2} \hat N_2)t} e^{-i\omega_{+} \hat b_+^{\dagger}\hat b_+t}e^{-i\omega_- \hat b_-^{\dagger}\hat b_-t} \\
& \quad \times e^{i(A_1\hat N_1 + A_2\hat N_2)}e^{i(B_1\hat N_1^2 +B_2\hat N_2^2)} e^{iD \hat N_1\hat N_2}  \\
& \quad \times \hat{D}_+[C_+ +K_{\hat{N}_+}\hat N_+]\hat{D}_-[C_-+K_{\hat{N}_-}\hat N_-] , 
\end{aligned}
\end{equation}
where,
\begin{equation}\label{Coefficients}
\begin{aligned}
\theta &= \theta_+ +\theta_-,\\
A_1&= \sqrt{2} \bar{g}_1(A_+c-A_-bs),\\
 A_2&=-\sqrt{2} \bar{g}_2\left(A_+\frac{s}{b} + A_-c\right),\\
B_1&= 2 \bar{g}_1^2(c^2B_+ +b^2s^2B_-), \\
B_2&= 2 \bar{g}_2^2\left(c^2B_- +\frac{s^2}{b^2}B_+\right), \\
D&= 4cs\bar{g}_1\bar{g}_2\left(bB_- - \frac{1}{b}B_+\right).
\end{aligned}
\end{equation}

When (\ref{eq:modulatedH}) are time-independent, then the solution (\ref{eq:Uk}) is well known \cite{Mancini1997,Bose1997,Armata2017, Qvarfort2018}. More recently, the extension to arbitrary time dependence has be been obtained in \cite{qvarfort2020time}. Here the solutions have the following integral representation (note the additional $\pm$ notation should not be confused with the transformed modes, $\mu=+,-$, appearing above),
\begin{equation}
\begin{aligned}
\theta_\mu &=-F_{\hat{B}_{\mu,+}} F_{\hat{B}_{\mu,-}}, \\
A_\mu&= -(F_{\hat{N}_\mu} + F_{\hat{B}_{\mu,+}}F_{\hat{N}_\mu\hat{B}_{\mu,-}} + F_{\hat{B}_{\mu,-}}F_{\hat{N}_\mu\hat{B}_{\mu,+}}), \\
B_\mu&=-(F_{\hat{N}_\mu^2} + F_{\hat{N}_\mu\hat{B}_{\mu,+}}F_{\hat{N}_\mu\hat{B}_{\mu,-}}),\\
C_\mu &=  F_{\hat{B}_{\mu,-}} -iF_{\hat{B}_{\mu,+}} ,\\
K_{\hat{N}_\mu} &= F_{\hat{N}_\mu\hat{B}_{\mu,-}}-iF_{\hat{N}_\mu\hat{B}_{\mu,+}} \,. 
\end{aligned}\label{UkCoefficients}
\end{equation} 
where (see \cite{qvarfort2020time} for details),
\begin{equation}
\begin{aligned}\label{eq:F:coeffs}
F_{\hat{N}_\mu}=& -2  \omega_\mu^2 \int_0^t dt' \mathcal{D}_{1,\mu}(t') \sin(\omega_m t')  \\
& \times \int_0^{t'}dt'' k_\mu( t'') \cos(\omega_\mu t'')  \\
& -  2  \omega_\mu^2 \int_0^t dt' k_\mu( t') \sin(\omega_m t')  \\
& \times \int_0^{t'}dt'' \mathcal{D}_{1,\mu}(t'') \cos(\omega_\mu t''),   \\
F_{\hat{N}^2_\mu} =&   - 2  \omega_\mu^2 \int_0^t dt' k_\mu( t')\sin(\omega_\mu t'), \\ 
& \times \int_0^{t'}dt'' k_\mu( t'') \cos(\omega_\mu t'') , \\
F_{\hat{B}_{\mu,+}}=&-\omega_{\mu} \int_0^t dt' \mathcal{D}_{1,\mu}(t')\cos(\omega_\mu t'), \\
F_{\hat{B}_{\mu,-}}=&  -\omega_{\mu} \int_0^t dt' \mathcal{D}_{1,\mu}(t')\sin(\omega_\mu t'),  \\
F_{\hat{N}_\mu\hat{B}_{\mu,+}}=& 
-\omega_\mu\int_0^t\,dt' k_\mu(t')\cos(\omega_\mu t'), \\
F_{\hat{N}_\mu\hat{B}_{\mu,-}}=& 
-\omega_\mu\int_0^t dt' k_\mu(t')\sin(\omega_\mu t') 
.
\end{aligned}
\end{equation}
For our purposes, the most the relevant terms are contained in the field-field interaction $\hat{U}_{12}(t)=e^{iD(t) \hat{N}_1\hat{N}_2}$, as well as the coefficients $K_{\hat{N}_\mu}$, which determine the coupling of the fields to the mechanics. The latter leads to a drop in purity of the combined optical state which in turn reduces the observable entanglement. While these terms in general will remain finite at all times $t>0$, for certain choices at least one $\mu$-mode can be made to decouple. For example, in the case of a single time-independent optomechanical system, a well known feature is that the field and mechanics disentangle at multiples of the mechanical period, $t=2\pi/\omega_m$. In our model, this corresponds to,
\begin{equation}\label{Kconst}
K_{\hat{N}_\mu}= -k_\mu(1-e^{-i\omega_\mu t}), 
\end{equation}
however, a non-zero gravitational interaction means, $\omega_+ \neq \omega_-$, and so both cavity fields cannot simultaneously decouple\footnote{Except at times when some multiple of the ratio of frequencies is itself an integer.}. 

In principle this means that the entanglement generated between the fields will always have some dependence on the initial states of the mechanical elements. We will asses the effect of this contribution in section \ref{sec:verification}, however a simple estimate of the entanglement rate can be made by temporarily ignoring the field-mechanics interaction. The evolution operator (\ref{Ugeneral}) then has a reduced part, 
\begin{equation}
\hat{U}_\text{red}=\hat{U}_1\hat{U}_2\hat{U}_{12}, \quad \text{where} \quad \hat{U}_{12}(t)=e^{iD(t) \hat{N}_1\hat{N}_2},
\end{equation}
acting only on the physical cavity field modes. For interactions of this form, we can show that the linear entropy, $S_L=1-\tr[\rho_1^2]$, for an initially separable pure state is given by,
	\begin{equation}\label{SLapprox}
	S_L = 2D(t)^2 (\Delta \hat{N}_1)^2 (\Delta \hat{N}_2)^2 + \hdots
	\end{equation}
(see appendix \ref{App:LinearEntropy} for details and the full expression).

\begin{figure}[htb]
\includegraphics[width=8cm,angle=0]{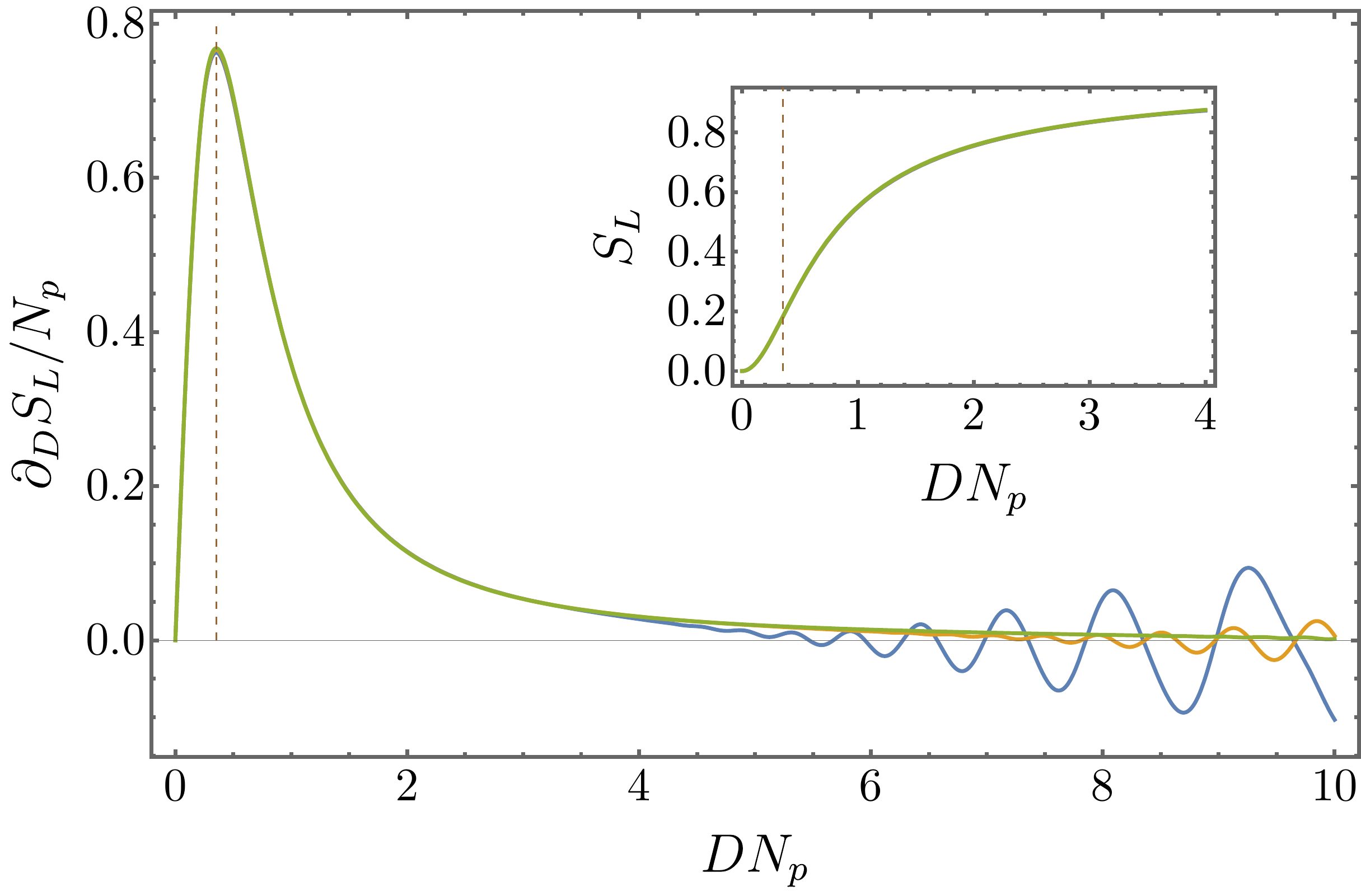}
\caption{\label{fig:SL} The gradient of the linear entropy (divided by the photon number), plotted as a function of $D N_p$ for $N_p=10$ (blue), $N_p=20$ (orange) and $N_p=40$ (green). For $S_L \ll1$, all three curves are almost identical (see inset plot). The brown dashed line corresponds to $D=\frac{1}{2\sqrt{2}N_p}$. Oscillations for large $D N_p$ (visible for the blue curve) are a consequence of periodicities in the linear entropy, which begin to appear around the point it reaches its maximum. All plots we made using (\ref{App:SLnumerics}). }
\end{figure}

A characteristic timescale, $\tau_e$, can then be estimated by considering the lowest order contributions to the linear entropy. Ignoring the numerical factor in (\ref{SLapprox}), we will take this as the solution to,
\begin{equation}\label{SL}
|D(\tau_e)| \propto\frac{1}{ \Delta \hat{N}_1 \Delta \hat{N}_2}.
\end{equation}
Thus, the entanglement rate, at least when $S_L$ is small, is enhanced by states with large initial variances. For fixed photon numbers, $N_p \equiv \avg{\hat{N}_j}$, the optimal choice is a superposition of Fock states, however generating these for large $N_p$ is difficult in optomechanical systems. A much higher photon number variance can be achieved by using coherent states, where $(\Delta \hat{N}_j)^2 = N_p$ (assuming equal photon numbers in each cavity). In this case, we observe that the general form (\ref{SL}) approximately holds for $S_L$ approaching one, where the proportionality factor must be determined numerically. A conservative choice is to pick the point of maximum rate of change with $D$, which corresponds to $|D(\tau_e)|\approx \frac{1}{2\sqrt{2}N_p}$, see figure \ref{fig:SL}. 

%--------------------------------------------------------------
\section{Estimation of Entanglement Time: Constant vs. Modulated Coupling}\label{Sec:EntanglementTime}
For (\ref{SLapprox}) to be valid, one needs to evaluate $D(t)$ at a time when both the $K_\mu(t)$ coefficients are sufficiently small. If the bare mechanical frequencies differ significantly, this may not be possible (we address this issue in section \ref{Sec:frequency_mismatch}), and in practice the choice will largely depend on whether one mode is easier to cool. For reasonably symmetric cavities, a sensible approach is to measure at a multiple of the average of the mode decoupling times. In the example above, this corresponds to $t_q= q\pi  (\frac{1}{\omega_+}+\frac{1}{\omega_-})$, where $q\in \mathbb{Z}^+$. However, in general we will adopt the label $t_q$ to correspond to the $q$th (chronologically occurring) optimal measurement time for a given strategy.
\begin{table*}[t!]
    		\centering
			\begin{tabular}{Sc Sc Sl} 
				\multicolumn{3}{Sc}{\textbf{Recurring functions/derived quantities}}\\ \hline
				  \textbf{Symbol}  &  \textbf{First appearance} &   \textbf{Meaning} \\ 
				\hline \hline
				   $D$ & Eq.\eqref{Ugeneral} & controls non-separability of cavity states \\
				    $K_{\hat{N}_\mu}$ &  Eq.\eqref{Ugeneral}  &  controls non-separability of states of cavity and mechanics  \\
				    $t_q$ &  above section \eqref{subsec:timeindsys} &  $q$th (chronologically occurring) optimal measurement time \\
                    $\tau_e$ &  Eq.\eqref{SL} &  entanglement time\\
                    $\kappa_\mu$ & Eq.\eqref{maintextWitSimple} & decoherence rates \\
                    $\delta t$ &  above Eq.\eqref{eq:Btq} & size of measurement window \\
                    $\mathcal{W}_1$ & Eq.\eqref{maintextWit1} & entanglement witness \\
                    $\gamma_g$  & below Eq.\eqref{frequencies} &  gravitational coupling parameter 
			\end{tabular} 
		\caption{  Table of the most important recurring functions and derived quantities.
			\label{tab:functions} }
	\end{table*}

\subsection{Time Independent Systems}\label{subsec:timeindsys}
In the case of time-independent optomechanical systems, it is straightforward to calculate the relevant coefficient in (\ref{eq:Uk}) necessary for estimating the entanglement rate. This is given by \cite{Mancini1997,Bose1997},
\begin{equation}\label{Bconstant}
B_\mu= k_\mu^2(\omega_\mu t-\sin \omega_\mu t).
\end{equation}
In the ideal scenario, the mechanical frequencies should be equal, $\omega_{m,j}\equiv \omega_m$, as this maximises the viable measurement window. We then note from (\ref{frequencies}), that the difference in mode decoupling periods is approximately $\Delta t=4 \pi \gamma_g/\omega_m$. Therefore, provided the number of mechanical periods $q \ll 1/\gamma_g$, then $D(t)$ will vary slowly between the respective decoupling times. Inserting the above into (\ref{Coefficients}), we find to lowest order in $\gamma_g$,
\begin{equation}
\begin{aligned}\label{eq:Dconst}
D(t_q)&= \frac{6G\sqrt{m_1m_2}k_1k_2t_q}{\omega_m d^3}, \qquad \frac{2 \pi q}{\omega_-} < t_q <\frac{2 \pi q}{\omega_+},
\end{aligned}
\end{equation}
where $k_j=\frac{ x_{0,j}g_j}{\hbar \omega_m}$ are the original coupling constants for each cavity. From (\ref{SL}) we can then estimate that, for coherent states, the characteristic entangling time is given by, 
\begin{equation}
\tau_e = \frac{\omega_m d^3}{6\sqrt{2}Gk_1k_2 \sqrt{m_1m_2}N_p}.
\end{equation} 
In general, achieving coherence for large masses on this timescale is a significant challenge and so the most natural approach is to choose $m_1=m_2\equiv m$ for the largest viable $m$. Realistic parameter values are given in Table \ref{tab:parameter:values}, then assuming $k_1=k_2 \equiv k_0$, suggests an entanglement timescale on the order of $\tau_e \sim 10^6\,$s.

\renewcommand{\arraystretch}{1.2} 
	\begin{table*}[t!]
		\centering
			\begin{tabular}{Sl Sc Sc} 
				\multicolumn{3}{Sc}{\textbf{Example parameters}}\\ \hline
				\textbf{Parameter} & \textbf{Symbol}  &  \textbf{Value} \\ 
				\hline \hline
				Separation & $d$ &  $2\times 10^{-4}$m  \\
				Oscillator density &$\varrho$& $2$ $ \mathrm{g \,cm^{-3}}$ \\
				Oscillator masses & $m $ &  $ 10\,\rm{ng}$  \\
				Oscillator radius & $ R$ &   $\sim 11\,\rm{\mu m}$   \\ 
				Mechanical frequency & $\omega_m$ & $2 \pi \times 10^3 $ rad s$^{-1}$\\
				Photon number & $N_p$ & $10^7$ \\ 
				Optomechanical mode coupling & $k_0$ & $1$  \\
				Environmental temperature & $T$ & $1\,$mK
			\end{tabular} 
		\caption{Hypothetical experimental parameters: $\varrho$ and $\omega_m$ are similar to those achieved in \cite{Matsumoto2019}, where a low-noise optomechanical system based with a silica mass in the mg range was implemented. $R$ is calculated assuming spherical mechanical elements. For these values $\gamma_{g}\approx 2\times 10^{-18}$. 
			\label{tab:parameter:values} }
	\end{table*}

Thus we see that even with high numbers of photons, the entanglement time is still long in comparison to typical noise timescales. 
A key result of the present work is that $\tau_{e}$ can be further decreased by modulating the parameters of the system. In particular, we will now show that driving the optomechanical coupling at, or close to, resonance with the frequency $\omega_m$ can lead to cubic time dependence for $D(t)$ in the long time limit. This leads to a significant speedup in the generation of correlations between the optical modes.

\subsection{Entanglement Enhancement Through Modulated Optomechanical Coupling}\label{ModulatedEntanglementTime}

We now assume that the light matter couplings are modulated according to $g_j(t)=g_0 + \bar{\epsilon} \cos (\Omega t + \phi_1) $. When resonant with one of the mode frequencies, this continually drives the (expected) oscillation amplitude, and position variance, of that mode \cite{Qvarfort2021}. As a result, entanglement generation is increased through the field-field channel. However, unlike in the constant coupling scenario, the mechanical state will vary after each oscillation and so no times $t>0$ exist when the field and mechanics factorise, i.e $K_\mu(t>0) \neq 0$.

The solution is to instead modulate the coupling at specific fractional frequencies of the form $\Omega_{n,\mu}=(1-1/n)\omega_{\mu}$, where $n \in \mathbb{Z}$. It was shown in \cite{Qvarfort2021} that for times $t_{q,\mu}=2 n q \pi/\omega_{\mu}$ the cavity and mechanical states do indeed disentangle. Evaluating the $F_{\hat{N}_\mu^2}$, and $F_{\hat{N}_\mu\hat{B}_{\mu,\pm}}$ integrals in (\ref{eq:F:coeffs}) we find that for an optimal phase choice, $\phi_1 =\pi/2$, and for times $t_{q,-}<t_q<t_{q,+}$, the $D$ coefficient is given by,
\begin{align}\label{eq:Dfrac}
 D(t_q) \approx 4 n q\pi \gamma_g \left(3 k_0^2 + \epsilon^2\frac{n^2(1 + (2n-1)^2)}{2(2n-1)^2}\right) ,
\end{align}
to first order in $\gamma_g=Gm/\omega_m^2 d^3$, where $k_0=x_0 g_0/(\hbar\omega_m)$ and $\epsilon=x_0 \bar{\epsilon}/(\hbar\omega_m)$. Note to this order, the choice between $\Omega_{n,\pm}$ is unimportant here. 

While $D(t_q)$ grows linearly in $q$, the term corresponding to modulated optomechanical coupling grows cubically in $n$. This suggests the most effective strategy is to measure at the first optimal measurement time ($q=1$). It is then clear that for $\omega_m t_1 \gg 1$, the second term will dominate (provided $\epsilon$ is not much smaller than $k_0$), that is, when multiple mechanical periods are needed to entangle the systems. For optomechanical cavities $\omega_m$ is typically larger than $2 \pi \times 10 \,\text{rad} s^{-1}$, and so if entanglement can be reached within $t\sim 10^{-2}s$, one would choose to work with unmodulated systems. The reality however, is that even for optimistic parameter choices, $ k_0^2 \gamma_g N_p \ll 1$, and so $\omega_m t_1$ needs to be large. This leads to our main result: in the long time limit, $|D(t_1)|\approx \frac{1}{4\pi^2} \epsilon^2 \omega_m^3 \gamma_{g}t_1^3$, and so the entanglement time for cavities initialised in coherent states is given by,
\begin{equation}\label{EntanglementRateModulatedg}
\tau_e \approx  \left( \frac{ 2\sqrt{2}\pi^2 d^3}{N_p \omega_m Gm  \epsilon^2 }  \right)^{\frac{1}{3}} .
\end{equation}
By way of comparison, using the parameters in Table \ref{tab:parameter:values} and setting $\epsilon= k_0 $, modulation decreases the entanglement time significantly to $\tau_e \approx 2\,$s, which corresponds to an improvement by $6$ orders of magnitude. In general, the modulated entanglement rate is enhanced by a factor of $\frac{1}{12}(\epsilon/k_0^3)^{2/3}(\pi N_p\gamma_g)^{-2/3}$ compared to the constant coupling case. This means it is particularly advantageous in precisely the experimental regimes which are most realistic: weak gravitational interaction, small light-matter coupling and low photon numbers.

In practice, the restriction to fractional frequencies does not significantly decrease the direct field-field coupling over the resonant case. Similar calculations for, e.g. $\Omega=\Omega_{\infty,+} = \omega_m$, lead to,
\begin{align}
\label{eq:Dcoeffmod}\nonumber
D(t_q)\approx & \,  \Bigg(6 k_0^2    - \frac{3}{4}\epsilon \left(  4   k_0  \cos  \phi_1 + \epsilon \cos 2 \phi_1\right) \\ 
& \quad + \frac{\epsilon^2}{2} \Bigg)\gamma_g \omega_mt_q   - \frac{1}{12} \gamma_g  \epsilon^2 \omega_m^3 t_q^3 \,.
\end{align}
where we have taken $t_q=2q\pi/\omega_m$. For $\omega_m t_q \gg 1$, the last term in equation (\ref{eq:Dcoeffmod}) dominates and so compared to (\ref{eq:Dfrac}) at a given target time (choosing the driving frequency as $\Omega_{{n=q},\pm}$) we see that the modulus differs by only a factor $\pi^2/3 \sim 3$.  On the other hand, the larger $K_{N_\mu}(2q\pi/\omega_m)$ values resulting from the resonant coupling impose stricter cooling requirements for the mechanical modes.

%---------------------------------
\section{Entanglement Verification and Noise}\label{sec:verification}
%---------------------------------
The results of the previous sections correspond to the best case entanglement rate between the two cavity fields. In reality, a combination of external decoherence along with other intra-system couplings will mean that the final field-field state will not be pure. In general, quantification (and measurement) of entanglement in mixed states is typically challenging, though there are notable exceptions -- as an example, in Appendix \ref{App:RhoCavity} we analyse the logarithmic negativity for initial cavity states of the form $\ket{\psi}=\frac{1}{\sqrt{2}}(\ket{0}+\ket{N})$, recovering the qualitative features outlined in the following sections. Instead it is often more practical to resort to so-called entanglement witnesses, $\hat{\mathcal{W}}$, which 
verify with certainty the presence of entanglement if $\mathcal{W}=\Tr \hat{\mathcal{W}} \rho <0$, but can make no statement otherwise \cite{Guhne2009}. 

From a technical perspective, this means one must be careful when using such operators to infer an entanglement rate. In particular, $\Tr \hat{\mathcal{W}} \rho(t_2)< \Tr \hat{\mathcal{W}} \rho(t_1)$ does not guarantee that entanglement has increased in time. On the other hand, achieving $\mathcal{W}(t)<0$ to begin with may impose a minimum constraint on the allowable noise in the system, and this \textit{can} have a meaningful time dependent operational interpretation. We will quantify this in the sections below. 

In practice, however, one will also need a sufficiently high signal-to-noise ratio\footnote{ Here we are referring to noise on the measurement of the optical state after leaving the system (e.g. from the detector), and not on decoherence effecting the state during the build up of entanglement. The latter will be addressed below.} in the measurement of $\mathcal{W}$ on the state for entanglement to be observable. 
The particular threshold needed (i.e. $\mathcal{W}< -\text{thresh.}$) will depend heavily on the experimental apparatus used. As such, identifying this cut-off \textit{a priori} is difficult, but we can maximise the chances of verification by performing the measurement when the witness function reaches its minimum value below zero.

In the following section we identify a witness that not only detects entanglement in our system, but also recovers approximately the same timescale (\ref{EntanglementRateModulatedg}) in the limit that all field-mechanics interactions vanish (at the chosen measurement time). More generally, this scaling is preserved as long as the noise constraints (which we will derive explicitly) are satisfied to within roughly an order of magnitude. This holds both in the case of extraneous intra-system couplings and for thermal decoherence.

%----------------------------------------
\subsection{Entanglement Witness}\label{Sec:Witness}

In order to identify an appropriate entanglement witness it is useful to anticipate the ideal joint state of the field modes. Intuitively, when the gravitational coupling strength is large, then the mirrors effectively become rigidly connected and so the system is analogous to a single cavity containing two modes \cite{Wan2017}\footnote{Note that this corresponds $\omega_s\rightarrow 0$.}. For certain parameter choices, it is known that these modes can become entangled, forming the two-mode cat state, $\ket{\psi}_{cav}=(1+i)\ket{\alpha_1}_1\ket{\alpha_2}_2+(1-i)\ket{-\alpha_1}_1\ket{-\alpha_2}_2$ \cite{Bose1997}. However, many of the simplest continuous variable entanglement criteria fail to witness entanglement in these types of states. One of the first examples that does was provided by Shchuckin and Vogel \cite{ShchukinVogel2005}, and can be represented by the following determinant,
\begin{equation}\label{maintextWit1}
\mathcal{W}_1(t) = \begin{vmatrix}
1 & \braket{a_2^{\dagger}} & \braket{a_1a_2^{\dagger}}  \\ 
\braket{a_2} & \braket{a_2^{\dagger}a_2} & \braket{a_1a_2^{\dagger}a_2}  \\ 
\braket{a_1^{\dagger}a_2} & \braket{a_1^{\dagger}a_2^{\dagger}a_2} &\braket{a_1^{\dagger}a_1a_2^{\dagger}a_2}  
\end{vmatrix}.
\end{equation}
This can be measured, in principle, using homodyne correlation measurements \cite{Shchukin2005a}.

In Appendix \ref{App:Witness} we provide explicit evaluations of the time dependent terms appearing in (\ref{maintextWit1}) assuming an initial state of the form $\rho(0)=\ketbra{\alpha_1}{\alpha_1}\otimes \ketbra{\alpha_2}{\alpha_2} \otimes \rho_{+}^{\text{th}}\otimes \rho_{-}^{\text{th}}$, where $\ket{\alpha}$ is a coherent state and $\rho_\mu^{\rm{th}}$ are thermal states of the mechanical modes. For simplicity, we will continue to work almost exclusively in the symmetric cavity approximation (though the extension to arbitrary parameters will turn out to be straightforward). Under these conditions $\cos(a)=\sin(a) =\frac{1}{\sqrt{2}}$ and $b=1$. The minus and plus mode (see Eq.\eqref{symplectictransforms}) then reduce to the simpler stretch and center of mass mode, respectively,
\begin{equation}
\begin{aligned}
    \hat x_s  &\equiv  \frac{\hat x_2 - \hat x_1}{\sqrt{2}} = \hat x_-\left(\cos(a)=\sin(a)=\frac{1}{\sqrt{2}},b=1\right), \\
    \hat{x}_c &\equiv \frac{\hat x_2 + \hat x_1}{\sqrt{2}} = \hat x_+\left(\cos(a)=\sin(a) =\frac{1}{\sqrt{2}},b=1\right),
\end{aligned}
\end{equation}
along with the associated number operators $\hat N_s = -(\hat N_1 + \hat N_2)$ and 
$\hat N_c = \hat N_1 - \hat N_2$.  
Similarly, $\mathcal{D}_{1,c}=0$, which implies $\theta_c=A_c=C_c=0$, and so adopting the same labelling convention, equations (\ref{Coefficients}) simplify to, 
\begin{equation}
\begin{aligned}\label{eq:CoefficientsRed}
\theta &=  \theta_s, \\
A_j &= -A_s \equiv A, \\
B_j& = B_c+B_s \equiv B, \\
D&= 2(B_s-B_c).
\end{aligned}
\end{equation}

In general, the resulting expectation values are functions of the absolute values, $|\alpha_1|$ and $|\alpha_2|$ of the coherent state parameters for the two cavity modes, and not the relative phase. In the symmetric cavity scenario a natural choice is to set these equal, $\alpha_1=\alpha_2=\alpha$, with $|\alpha|^2=N_p$. Even then, the full witness expression is still difficult to treat analytically, however, numerical results suggest that the restriction to $B=0$ at the measurement time provides a somewhat optimal regime. While this term has no effect on the entanglement, similar terms are known to influence the sensitivity of optomechanical sensors under certain POVMs (e.g. homodyne measurements, \cite{Qvarfort2021}). For $\epsilon=0$, it is usually sufficient to consider only integer values of $k_0$, but more care must be taken in the modulated scenario. In these cases, the witness 
is given by the simpler expression (\ref{appWit1noise}),
\begin{equation}\label{maintextWitSimple}
\begin{aligned}
	\mathcal{W}_1(t) & = N_p^3 \Big( 1 - e^{-4\kappa_c -4N_p(1-\cos D)}  \\
	 &\quad  - 2 e^{-\kappa_c-\kappa_s - 2N_p (1-\cos D)} \\
	 &\quad \times \left(1-  e^{-2\kappa_c - 2N_p(1-\cos D)}\cos D\right) \Big),
\end{aligned}
\end{equation}
which holds both when $\kappa_\mu$ describes the effects of the traced out thermal mechanical modes and, as we will see later, decoherence from an environment. It is therefore convenient to write $\kappa_{\mu}=\kappa_\mu^\mathrm{th}+\kappa_\mu^\mathrm{dec}$, where in the present section we will focus on the first term, given by $\kappa_\mu^\mathrm{th} = |K_{\hat{N}_\mu}|^2(\bar{n}_\mu+1/2)$, where $\bar{n}_\mu=1/(e^{\hbar\omega_s/k_B T_\mu}-1)$ is the occupation number for temperature $T_\mu$ (see (\ref{BetaAvgsthermal})).

In general, the optimal measurement time will lie somewhere between what would be the individual mode decouplings. For equal temperatures, this turns out to be just the mean, $t_q=q n\pi(1/\omega_c+1/\omega_s)$, where to preserve symmetry we will from now on choose the modulation frequency as $\Omega_n=\frac{1}{2}(1-\frac{1}{n})(\omega_c+\omega_s)$. To lowest order in $\gamma_g$, and for large $n$, both $\kappa^{\rm{th}}_\mu(t_q)$ are equal to (\ref{App:Kapprox}),
\begin{equation}\label{Kapproxlargen}
\kappa^{\rm{th}} (t_q)\approx \frac{1}{2}(\pi q n \gamma_g)^2  \left( 4 k_0^2 +  n^2 \epsilon^2\right)(\bar{n}+1/2).
\end{equation}
Neglecting the decoherence contribution, we can then set $\kappa_{\mu}=\kappa$ in the witness (\ref{maintextWitSimple}), which further simplifies to 
\begin{equation}\label{WitSimple2}
\mathcal{W}_1(t_q)= 4N_p^3 e^{-4y}\left(e^{2y}\sinh^2(y)-z^2\right),
\end{equation}
where $y=2N_pz^2+\kappa(t_q)$ and $z=\sin(D(t_q)/2)$. 

We can now apply the requirement that $\mathcal{W}_1(t_q)<0$ to give an upper bound on the thermal noise term. This leads to, 
\begin{equation}\label{kappabound}
\kappa(t_q) <  \frac{1}{2}\ln{(1 \pm 2 z)} -2N_pz^2.
\end{equation}
If an experiment is sensitive to small amounts of entanglement, such that very early measurements are viable ($D(t_q) \ll 1/N_p$) then we can expand the right hand side to first order in $z$,
\begin{equation}\label{kappalimit}
\kappa(t_q)\lesssim  |z| \approx \frac{1}{2}|D(t_q)|.
\end{equation}
This gives the highest acceptable noise for which entanglement can be verified. We find a similar condition for the logarithmic negativity when the initial cavity fields are in superpositions of Fock states (see (\ref{eq:LNtapp})), suggesting the behaviour of the witness is reasonably faithful for small amounts of entanglement. In particular, for $\kappa=0$, entanglement can be immediately detected for any non-zero $D(t_q)$. 

On its own, however, this is not especially useful, at least in the case of purely thermal noise. This is because $\kappa^{\rm{th}}(t_q)$ and $D(t_q)$ have different time dependencies, and indeed the former tends to grow faster than the field-field coupling. As a result, (\ref{kappalimit}) only provides an upper bound on when entanglement can be detected. To see this explicitly, it is convenient to separately consider the constant and modulated-dominant regimes. For later, we will also find it useful to regard $\kappa$ in general as a function of $z\approx D(t_q)/2$, then using (\ref{Kapproxlargen}) and (\ref{eq:Dfrac}) we have
\begin{equation}\label{kappathermalz}
\kappa^{\rm{th}}(t_q) \approx \begin{cases}
\frac{2\bar{n} +1}{36k_0^2}z^2 ,& \epsilon=0 \\
\frac{1}{2}(\bar{n} +\frac{1}{2})\left(\frac{\pi q \gamma_g}{\epsilon}\right)^{\frac{2}{3}}z^{\frac{4}{3}}, & \epsilon\gg \frac{k_0}{n},\, n\gg 1
\end{cases} 
\end{equation}
to lowest order in $\gamma_g$. Applying the inequality (\ref{kappalimit}), and inverting for $t_q$ we find that in both cases (up to a small numerical difference), $t_q \lesssim 12/(\omega_m \gamma_g (2 \bar{n} +1))$. Thus, thermal noise limits how long an experiment can be performed. In fact, a finite measurement time is a generic feature of our chosen witness -- even for $\kappa=0$, the next order expansion of (\ref{kappabound}) leads to $|D_{\max}(t_q)|\sim 2 |z| \lesssim \frac{2}{1+2N_p}$, despite the fact that the maximum entanglement has not yet been reached (see figure \ref{fig:SL}).

A more useful indicator for the ideal measurement time is to instead look for the point of maximum witness violation. This can seen in figure \ref{fig:witness}, where $\mathcal{W}_1(t_q)$ reaches a minimum of $-\frac{N_p}{4}$ in the limit $\kappa^{\rm{th}}\rightarrow 0$. We can then use this as a reference point to determine the maximum thermal noise for which the witness only drops to some fraction of this value.

\subsection{Optimal Measurement Time}
To proceed, it will be convenient to take $\kappa=\kappa(z)$, then the minimum of (\ref{WitSimple2}) is achieved for $z_{\min}$ satisfying, 
\begin{equation}\label{WitnessTPcondition0}
4z_{\min}^2+e^{2y_{\min}}=1+\frac{2z_{\min}}{4N_pz_{\min}+\kappa'|_{z_{\min}}}.
\end{equation}
If $\kappa$ varies slowly with respect to $z$ (i.e. $\kappa'(z)\ll 4N_pz$) we can ignore the noise gradient term in the denominator. Substituting for $y_{\min}$ we then have, 
\begin{equation}\label{WitnessTPcondition}
4N_pz_{\min}^2=\ln\left(1+ \frac{1}{2N_p}-4z_{\min}^2\right) -2\kappa(z_{\min}).
\end{equation}
Now, for realistic values of $\gamma_g$, $z_{\min}$ must be small on accessible timescales. This suggests $N_p \gg 1$ (which we will assume from now on), and so expanding the logarithm to first order, we have after a simple rearrangement\footnote{Alternatively (\ref{WitnessTPcondition}) can be formally solved in terms of the Lambert-W function. An asymptotic expansion in the large $N_p$  limit recovers the same result without explicitly assuming small $z$.}, 
\begin{equation}
D(t_q) \approx 2z_{\min}\approx  \frac{\sqrt{1-4N_p\kappa}}{\sqrt{2}N_p} .
\end{equation}
Thus, in the $\kappa\rightarrow 0$ limit, the witness will take its minimal value when $D(t_q)\approx 1/\sqrt{2}N_p$. Substituting into (\ref{WitSimple2}), we have,
\begin{equation}
\mathcal{W}_{1,\min}(t_q)|_{\kappa\rightarrow 0} \approx -\frac{N_p}{4},
\end{equation}
as claimed.

At high temperatures, the slowly varying condition is typically not satisfied, and so the derivative term in (\ref{WitnessTPcondition}) must be included. Following from (\ref{kappathermalz}), we can write $\kappa^{\rm{th}}=\Gamma_0^{\rm{th}}z^2$ and $\kappa^{\rm{th}}=\Gamma z^{4/3}$, for the constant and modulated optomechanical coupling regimes separately. The minimum conditions in the first case are easily found by noting that (\ref{WitnessTPcondition0}) is equivalent to (\ref{WitnessTPcondition}) with the substitutions $N_p\rightarrow N_p+\Gamma_0^{\rm{th}}/2$ and $\kappa\rightarrow0$. We then find, 
\begin{equation}\label{Dconstopt}
D(t_q) \approx\frac{1}{\sqrt{2}(N_p+\Gamma^{\rm{th}}_0/2)},
\end{equation}
which implies the measurement time should satisfy,
\begin{equation}
t_q \approx \frac{1}{6 \sqrt{2} k_0^2 \gamma_g \omega_m (N_p +\Gamma^{\rm{th}}_0/2)},
\end{equation}
(where we have again used $N_p\gg 1$). Thus, just as with $D_{\max}$, the optimal timescale shifts earlier at higher temperatures. This is, of course, not an advantage because the added noise leads to an increase in the minimum value of the witness (see figure \ref{fig:witness}). At the first optimal measurement time, this is given by,
\begin{equation}
\mathcal{W}_{1,\min}(t_1)\approx-\frac{N_p^{ 3}}{(2N_p+\Gamma^{\rm{th}}_0)^2}.
\end{equation}
In effect, it prevents the experiment from running long enough to build up a highly entangled state. We can characterise an acceptable level of noise by demanding that the magnitude of the witness decreases by no more than some factor, $a$, of the idealised ($\kappa^{\rm{th}}=0$) value. This amounts to solving $\mathcal{W}_{1,\min}=a\mathcal{W}_{1,\min}|_{\kappa\rightarrow 0}$, for which we find $\Gamma^{\rm{th}}_0 \lesssim 2\left(\frac{1}{\sqrt{a}}-1\right)N_p$.
Therefore, to keep the witness below, for example, half its absolute minimum (for a given set of experimental parameters), the mechanical modes must both be cooled to,
\begin{equation}\label{nbarmaxconst}
\bar{n}_{\max,\epsilon=0}<36k_0^2\left(\sqrt{2}-1\right)N_p-1/2.
\end{equation}
As $N_p$ will typically be very large, this should not pose a significant obstacle. 

For modulated optomechanical couplings, the analysis proceeds along similar lines though it is convenient to first make a suitable expansion of (\ref{WitnessTPcondition0}) (see appendix \ref{App:Witness} for details). In this case we find (to first order in $1/N_p$), 
\begin{equation}\label{Modulatedz}
	\begin{aligned}
		z_{\min}(t_1)&\approx\frac{1}{2\sqrt{2}N_p}\left(1-v\right)^{3/2}, \quad \text{and}\\
		t_1& \approx   \left( \frac{ 2\sqrt{2}\pi^2}{N_p \omega_m^3  \epsilon^2 \gamma_g}  \right)^{\frac{1}{3}} \sqrt{1-v},
	\end{aligned}
\end{equation}
where $v=\frac{5}{18}\frac{\bar{n} +\frac{1}{2}}{N_p^{1/3}}\left(\frac{\pi \gamma_g}{\epsilon}\right)^{\frac{2}{3}}$. It is immediately clear that for all but very large $\bar{n}$, the minimal witness time is very close to that for $\kappa=0$. Indeed, substituting into (\ref{WitSimple2}), and choosing $a=1/2$, eventually leads to the bound,
\begin{equation}\label{nmax_mod}
\bar{n}_{\max} \lesssim 0.66\left(\frac{ \epsilon}{\pi \gamma_g}\right)^{\frac{2}{3}} N_p^{1/3}- \frac{1}{2}.
\end{equation}
Thus, for realistic experimental parameters (where $\gamma_g \ll 1$) modulated couplings significantly improve tolerance to thermal noise. Indeed, for the values in table \ref{tab:parameter:values} and $\epsilon=k_0$, $\bar{n}\lesssim 4\times 10^{13}$, which corresponds to a mode temperature of $T\le  2\times 10^6\,$K.

%--------------------------------------------------------------
\subsubsection{Thermal Dependence of the Measurement Window}
The constraints above are valid when the measurements are localised to within a very short window around $t_q$ (which we can expect is less than the decoupling time difference, $\delta t \sim 2 \gamma_g t_q$). It is important, however, to know the extent of the usable measurement window, both because such processes are not instantaneous and because reducing the signal to noise ratio will rely on obtaining long integration times. The easiest way to estimate this is via the condition (\ref{kappabound}), along with expressions for $D(t_q\pm\delta t)$ and $\kappa^{\rm{th}}(t_q\pm\delta t)$.  

This approach will only be viable provided $B(t_1\pm \delta t)\approx 0$, otherwise the full witness expression must be used. In general, however, the window will be small on the timescale of the mean mechanical period $\bar{\tau}_m=\pi(1/\omega_c + 1/\omega_s)$. Expanding in terms of $\delta t= \zeta\bar\tau_m$ and assuming $\gamma_g{\ll}\zeta\ll1$, it can be shown that,
\begin{equation}\label{eq:Btq}
B(t_q+\zeta\bar{\tau}_m) \approx B(t_q)+\frac{4}{3}k_0^2 \pi^3\zeta^3,
\end{equation}
(where additional $\epsilon$ terms only enter at order $\gamma_g^2 \zeta^2$). Similarly we can find, 
\begin{equation}\label{kappawindow}
\kappa^{\rm{th}}(t_q+\zeta\bar{\tau}_m) \approx \kappa^{\rm{th}}(t_q) + 2k_0^2\pi^2(\bar{n} +1/2) \zeta^2,
\end{equation}
and,
\begin{equation}
D(t_q+\zeta\bar{\tau}_m) \approx D(t_q)-4qk_0\epsilon (n\pi )^2\gamma_g \zeta\,.
\end{equation}
The first of these expressions tells us that even at zero temperature, $B$ will be much smaller than $\kappa^{\rm{th}}$, and so we can reasonably expect (\ref{maintextWitSimple}) to hold. We also note that for $\kappa^{\rm{th}}$ the variation in $\zeta$ does not depend on $\gamma_g$, and so thermal noise grows much faster than the coupling $D$. This means the measurement window will be much smaller than $\bar{\tau}_m$ (as $D(t_q) \sim O[\gamma_g]$), and we therefore take $D$ to be effectively constant over this timescale. Moreover, when $\bar{n}\ll \bar{n}_{\max}$, we can also ignore the first term in (\ref{kappawindow}) and so around the decoupling time, $\kappa^{\rm{th}}$ is independent of $D$. This immediately allows us to determine both the conditions and size of the maximum measurement window: from (\ref{kappabound}), the maximum $\kappa^{\rm{th}}$ for which the witness is exactly zero occurs around $t_q$ satisfying, 
\begin{equation}
D(t_q)=2\sin^{-1}\left(-\frac{1}{4} \pm \frac{1}{4}\sqrt{1+\frac{2}{N_p}}\right),
\end{equation}
where for small $D$, we take the positive solution, and so, 
\begin{equation}\label{Doptimalwindow}
D(t_q) \approx \frac{1}{2N_p},
\end{equation}
with a corresponding $\kappa^{\rm{th}}_{\max}(t_q+\zeta\bar{\tau}_m) \approx \frac{1}{8N_p}$. Substituting in (\ref{kappawindow}) when $\bar{n}\ll \bar{n}_{\max}$, we find that the maximal size of the measurement window is\footnote{Thus within the window, $B(t)\ll D(t) \sim 1/N_p$, and so from (\ref{appWit1full}) we are justified in setting $B(t)=0$ in the analysis above.},
\begin{equation}\label{windowthermalopt}
\delta t \approx \frac{1}{2 \omega_m k_0 \sqrt{N_p(\bar{n}+\frac{1}{2})}}.
\end{equation}

At zero temperature, this equates to around $\pm 36$ns for the system outlined in table \ref{tab:parameter:values}, which is approaching the bandwidth limits of state of the art detectors. We note that in this case the window size is dominated by the constant component of the optomechanical coupling. 
If one can instead engineer $k_0=0$, $\epsilon>0$, then the next nonzero contribution to $\kappa^{\rm{th}}$ is $4$th order in $\zeta$,
\begin{equation}\label{kappawindow2}
\kappa^{\rm{th}}_{|_{k_0=0}}(t_q+\zeta\bar{\tau}_m) \approx \kappa^{\rm{th}}_{|_{k_0=0}}(t_q) + 2\epsilon^2\pi^4(\bar{n} +1/2) \zeta^4,
\end{equation}
and so the maximum $\delta t$ is instead given by,
\begin{equation}
\delta t_{|_{k_0=0}} \approx \frac{1}{\omega_m(N_p (\bar{n}+\frac{1}{2})\epsilon^2)^{1/4}}.
\end{equation}
By comparison, the zero temperature window is increased to $\pm3.4\mu$s, while making no change to the measurement time.
 
\begin{figure*}
		\subfloat[]{	\includegraphics[width=8cm,angle=0]{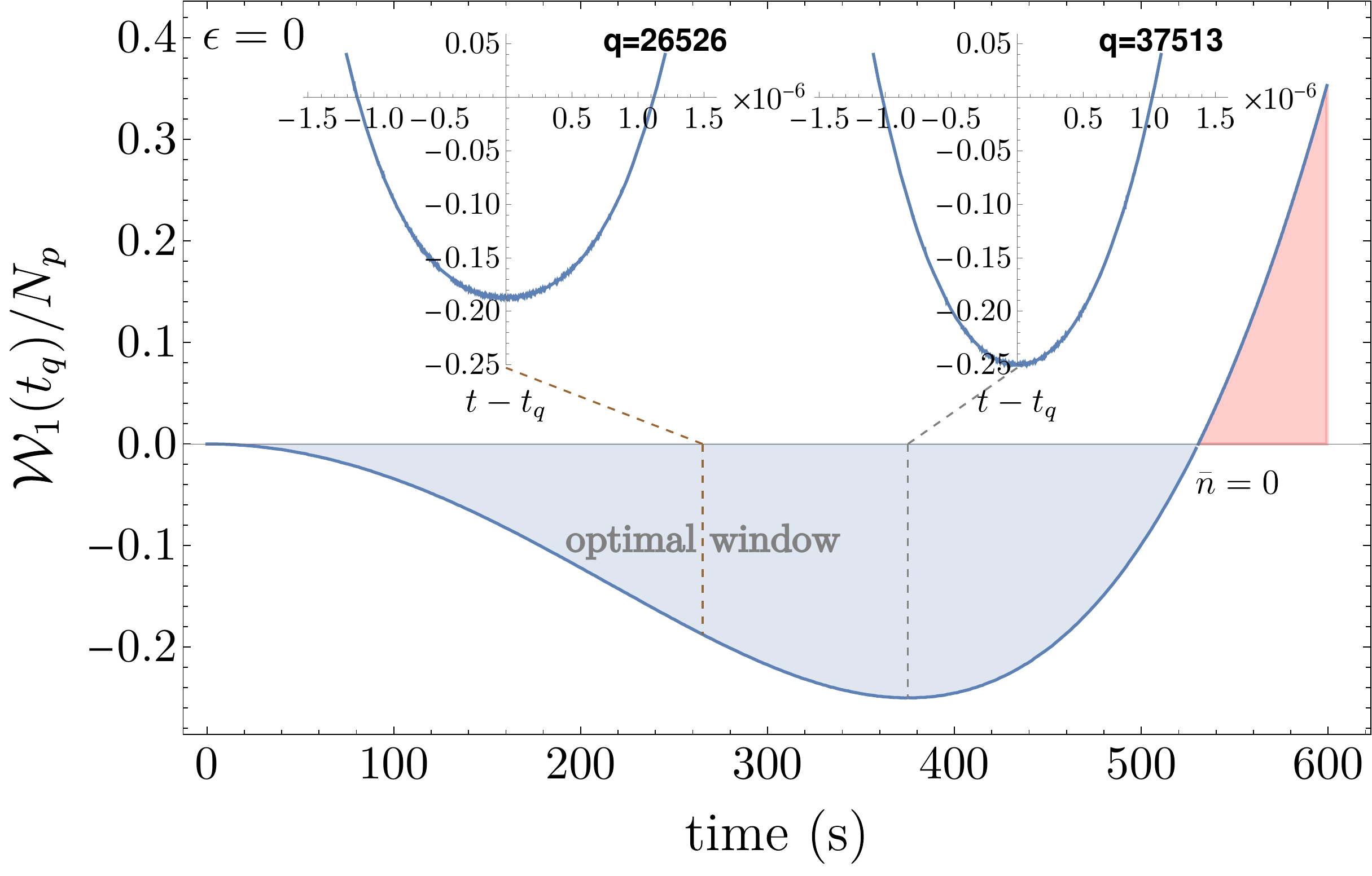}}
		\hspace{0.5cm}\subfloat[]{	\includegraphics[width=8cm,angle=0]{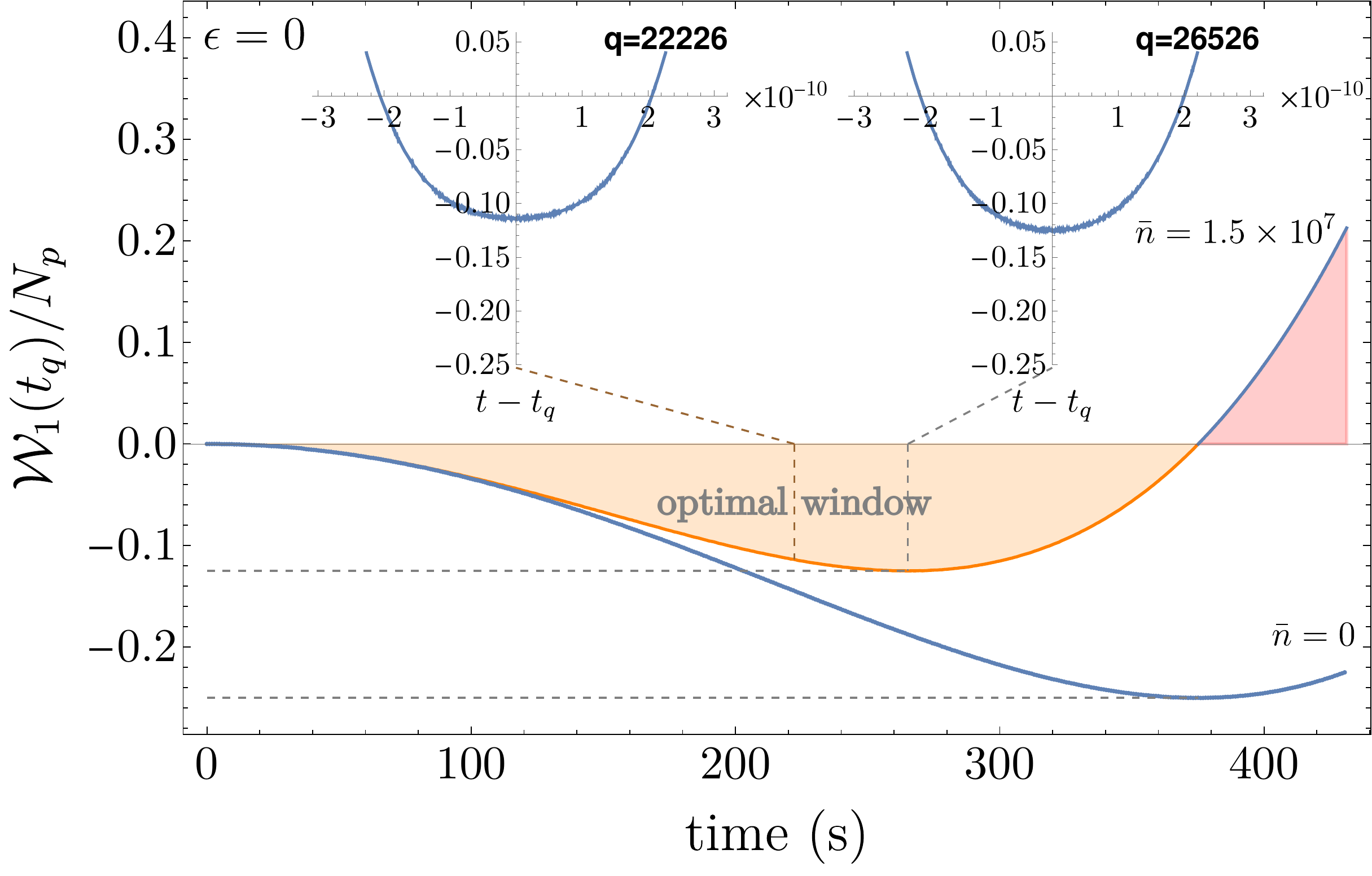}}
		\caption{The (rescaled) witness as a function of discrete $t_q$ times for constant optomechanical coupling ($n=1$), at both $\bar{n}=0$ ($a$) and the predicted $\bar{n}_{\max}$ given by (\ref{nbarmaxconst}) ($b$). Here we choose the more optimistic parameters, $\omega_m=2 \pi \times 10^2$Hz and $m=2.4\times 10^{-8}$kg, with $N_p=10^6$, leading to $\gamma_g = 5 \times 10^{-13}$. Between any two $t_q$'s, the witness can only verify entanglement within a small measurement window $|t-t_q| \lesssim \delta t $ (see insets). Numerical results agrees well with the theoretical value (\ref{windowthermal}), even beyond its apparent regime of validity. However, at high temperatures, the $t_q$ with maximal measurement window no longer follows (\ref{Doptimalwindow}). Note, there exists a maximum verification time (see (\ref{maxtime})), after which the witness is effectively always positive, even though entanglement may still be expected.  }\label{fig:witness}
\end{figure*}

More generally, we can estimate the measurement window at an arbitrary decoupling time by evaluating (\ref{kappabound}) using (\ref{kappawindow}) and $D(t_q)$. In the physically relevant regime, it is enough to expand (\ref{kappabound}) to second order in $D(t_q)$. We then have,
\begin{equation}\label{windowthermal}
\delta t \approx \sqrt{\frac{D(t_q)(1-N_pD(t_q))}{k_0^2 \omega_m^2(\bar{n}+1/2)}},
\end{equation}
where again we have taken $N_p \gg 1$. At the witness minimum, we typically have $D(t_q) \sim 1/\sqrt{2}N_p$, and so the window at this time is still around $91\%$ of its maximum value. Surprisingly, the estimates (\ref{windowthermalopt}) and (\ref{windowthermal}) appear to work well even at temperatures where the analysis breaks down, see figure \ref{fig:witness}.

%--------------------------------------------------------------
\subsubsection{Frequency mismatch}\label{Sec:frequency_mismatch}
The results of the previous sections are reasonably tolerant to most sources of asymmetry. However, the most obvious exception is when there are mismatches in the mechanical frequencies. This is because the separation between mode periods grows and so it becomes much harder to find a time when both $\kappa^{\rm{th}}_\mu$ are small. 

As the analysis in this case is significantly more complicated, we estimate the dependence only in the unmodulated scenario. Using the general frequencies $\omega_\mu$ (\ref{frequencies}), we insert the $B_\mu$ coefficients from (\ref{Bconstant}) into (\ref{Coefficients}) and expand $D(t)$ to lowest order in the (sum of the) relative frequency shifts in $\tilde{\omega}_\mu$, i.e. $\sum_j \frac{2\gamma}{m_j \omega_{m,j}}$. Assuming $\omega_{m,1} \geq \omega_{m,2}$ (wlog), we find after some work,
\begin{eqnarray}
D(t) &\approx&
\frac{4G k_1 k_2 \sqrt{m_1m_2}}{d^3\sqrt{\omega_{m,1}\omega_{m,2}}}\\
\nonumber &&\!\!\!\!\!\times \left( t + \frac{ \frac{\omega_{m,2}^2}{\omega_{m,1}} \sin(\omega_{m,1} t) - \frac{\omega_{m,1}^2}{\omega_{m,2}} \sin(\omega_{m,2} t) }{\omega_{m,1}^2-\omega_{m,2}^2 }\right),
\end{eqnarray}
where $k_j=\frac{ x_{0,j}g_0\bar{g}_j}{\hbar \omega_{m,j}}$, with $g(t)=g_0$, are the original coupling constants for each cavity. In deriving the above it is useful to adopt $k_+=(\frac{\omega_{m,1}}{\omega_+})^{3/2} k_1/(\sqrt{2}\bar{g}_1)$ and $k_-=(\frac{\omega_{m,2}}{\omega_+})^{3/2} k_2/(\sqrt{2}\bar{g}_2)$.

We now assume a small frequency mismatch, $\omega_{m,1}=(1+\xi_f)\omega_{m,2}$, where $\xi_f$ is a positive constant. Around a measurement time $t_q=\pi q(1/\omega_+ +1/\omega_-)$, we have,
\begin{equation}
D(t_q+\zeta\bar{\tau}_m)=D(t_q)(1-\xi_f)\left(1-\frac{2\pi^2 \zeta^2}{3}\right),
\end{equation}
where $D(t_q)$ is as given in (\ref{eq:Dconst}) (for $\omega_{m,2}=\omega_m$). Provided $\xi_f\ll 1$, we again see that $D(t)$ varies slowly around $t_q$. A similar analysis can be performed for $\kappa^{\rm{th}}_{\mu}$ terms, making use of (\ref{Kconst}). For $\xi_f$ much larger than the relative frequency shifts, we find
\begin{equation}
	\begin{aligned}
		\kappa^{\rm{th}}_+(t_q+\zeta\bar{\tau}_m)  & \approx \frac{k_1^2\pi^2}{2} \left(\bar{n}_+ + \frac{1}{2}\right)\\
		 & \quad\times (q^2 \xi_f^2 +4 q \zeta \xi_f + 4\zeta^2), \\
		\kappa^{\rm{th}}_-(t_q+\zeta\bar{\tau}_m)  & \approx \frac{k_2^2\pi^2}{2} \left(\bar{n}_- + \frac{1}{2}\right)\\
		& \quad \times (q^2 \xi_f^2 -4 q \zeta \xi_f + 4\zeta^2),
	\end{aligned}
\end{equation}
where again we are considering only the $\bar{n}_{\mu} \ll \bar{n}_{\max}$ limit so that $\kappa^{\rm{th}}_{\mu}(t_q)$ can be neglected in comparison to $D(t_q)$. Note that the expressions above do not depend on the gravitational field, which could have already anticipated from (\ref{kappawindow}). Thus $\bar{\tau}_m$ can effectively be taken as the mean of the bare mechanical periods, which makes the calculation significantly easier. For the sake of simplicity, we assume $k_j=k_0$, $\bar{n}_\mu=\bar{n}$ and $\zeta \sim \xi_f$ then in the long time limit (large $q$), $\kappa^{\rm{th}}_{+}\approx \kappa^{\rm{th}}_{-}$ is dominated by the $\xi_f$ term. Using the analysis of the previous section, we immediately find, 
\begin{equation}\label{freqboundopt}
	\xi_f \lesssim \frac{1}{2 \pi  q k_0 \sqrt{N_p(\bar{n}+\frac{1}{2})}}.
\end{equation}
Thus we see the trade-off between controlling thermal noise and the matching of mechanical frequencies -- even for perfectly localised measurement times, a non-zero $\Delta \omega_m$ can severely limit the upper bound on $\bar{n}$.

Note, the difficulties highlighted in the previous two sections are not inherent to optomechanical systems alone. Recall that the noise term $\kappa^{\rm{th}}(t)$ is directly proportional to $|K_{\hat{N}_\mu}(t)|^2$, which itself is related to the total displacement in phase space (as can be seen from (\ref{Ugeneral})). Thus the measurement window acts as a proxy for the level of control needed to return the mechanical degrees of freedom to close to their initial state.

%---------------------------------
\subsection{Thermal Decoherence}\label{Sec:Decoherence}

The relative insensitivity to the initial mechanical state arises because $K_{\hat{N}_s}(t_1)$ can still be very close to zero even at the center of mass mode decoupling time (or vice-versa). A more significant experimental challenge comes from external (open-system) noise processes, which in general cannot be avoided by choice of the measurement time. These can be broadly split into those acting directly on either the optical or mechanical degrees of freedom\footnote{Though for strong optomechanical couplings, there can be an interplay between these \cite{Naseem2018}. }. Optical losses, for example, can be expected to reduce the visibility of the signal \cite{Matsumura2020}, while mechanical decoherence will prevent the initial build up of entanglement between the masses. 

The latter in particular is well known for limiting the generation of macroscopic (spatial) superpositions. In a typical analysis, one modifies the usual von Neumann evolution equation by the inclusion of additional superoperators describing the effects of the environment (along with a corresponding renormalisation of the system Hamiltonian). If the temperature of the environment is assumed to be high (with an ohmic spectral density) then it is usually sufficient to consider only the decoherence term, $L[\rho]=-\frac{\Upsilon}{\hbar^2}[\hat{x},[\hat{x},\rho(t)]]$.  In this limit the coefficient is time independent, and given explicitly by $\Upsilon=2mk_BT \gamma_R$ where $\gamma_R$ is the relaxation rate and $T$ is the temperature of the environment \cite{Zurek2003}.
The extension to multiple mechanical modes is straightforward as long as the intra-system couplings are weak. In this case a good approximation to the dynamics can be found by adding additional $L[\rho]$ terms for each system degree of freedom \cite{Rivas2010}. Neglecting optical noise, our optomechanical model can then be described by the following equation,
\begin{equation}\label{maintextmastereq}
	\frac{{d \rho(t)}}{dt} = \frac{1}{i\hbar} [\hat{H}_{\text{ren}},\rho(t)] +L_1[\rho] +L_2[\rho],
\end{equation}
where $\hat{H}_{\text{ren}}$ is the renormalised system Hamiltonian (which for simplicity we will set equal to $H$, as this typically amounts to a shift in the mechanical frequencies), and $\gamma_R=\omega_m/Q$ is the relaxation rate for an optomechanical cavity with quality factor $Q$.

In order to evaluate the witness (\ref{maintextWit1}), we only need to solve for the reduced total optical state. For the single cavity (time-independent) setting, a convenient approach has been presented in \cite{Adler2005} where the analogue of the master equation (\ref{maintextmastereq}) is first projected out in terms of the operator $\rho_D(t)=\avg{p,q|\rho(t)|m,n}$, where $\ket{m,n}$ were Fock states of two arms of an interferometer. This can be solved exactly when making use of the partial trace over the mechanical modes, leading directly to the reduced field-field density matrix $\rho^{cav}_{p,q,m,n}(t)=\Tr_M[\rho_D(t)]$. 

The extension to two cavities is largely straightforward once we move to the decoupled mode picture, in particular where we note that for identical cavities, $L_1[\rho]+L_2[\rho]=L_c[\rho]+L_s[\rho]$. For completeness we provide full details in appendix \ref{App:Noise}, where an explicit solution is given by (\ref{App:rhocavsol}). This can be compared directly to the noiseless cavity reduced state (\ref{thermaldensity}), where we find that the modifications can be completely described by formally replacing the \textit{initial} thermal occupation numbers, $\bar{n}_\mu$, at time $t$, by
\begin{equation}
	\bar{n}_\mu(t)=\bar{n}_\mu(0)+\frac{\kappa_\mu^\mathrm{dec}(t)}{|K_{\hat{N}_\mu}|^2},
\end{equation}
where 
\begin{equation}\label{eq:kappadef}
		\kappa_\mu^\mathrm{dec}(t)= \frac{8\Upsilon}{\hbar^2}x_{0,\mu}^2 \int_0^t  \Re[e^{-i\omega_\mu (s-t)} K_{\hat{N}_\mu}(s-t)]^2 ds .
\end{equation}
As these terms always appear multiplied by an additional $|K_{\hat{N}_\mu}|^2$ factor,
(such that the thermal decay coefficient $\kappa_\mu \rightarrow \kappa_\mu^\mathrm{th}+\kappa_\mu^\mathrm{dec}$) any potential divergence in the second term is of no concern. Hence, while $K_{\hat{N}_{\mu}}(t)$ can in principle be zero, $\kappa_{\mu}^\mathrm{dec}(t)$ is not in general. We also stress that $\bar{n}_\mu(t)$ should not be interpreted as the current temperatures of the mechanical modes, but rather only as an effective model for the decoherence induced loss of purity of the cavity state (which occurs even for perfect decoupling).  

At zeroth order in $\gamma_g$, and for $\bar{n}\ll \bar{n}_{\max}$ we find that for identical cavities, $\kappa_{\mu}^\mathrm{dec}(t_q)= \kappa^\mathrm{dec}(t_q) \gg \kappa^\mathrm{th}(t_q)$, where
\begin{equation}\label{main:kappadec}
\begin{aligned}
\kappa^\mathrm{dec}(t_q) &= \frac{ 4\Upsilon}{\hbar^2}x_{0,m}^2 \frac{qn\pi}{\omega_m} \left(3 k_0^2 + \epsilon^2 \frac{n^2(1+(2n-1)^2) }{2(2n-1)^2}\right)\\
& \approx  \frac{ \Upsilon x_{0,m}^2}{\hbar^2\omega_m \gamma_g } D(t_q),
\end{aligned}
\end{equation}
Thus, from (\ref{kappalimit}) one immediately sees that entanglement can only be witnessed when,
\begin{equation}\label{decoherence_bound}
\gamma_R k_BT\lesssim \frac{\hbar G m}{2d ^3},
\end{equation}
which is independent of both the local dynamics and $N_p$. A similar behaviour is found for initial cavity states of the form $\ket{\psi}=\frac{1}{\sqrt{2}}\left(\ket{0}+\ket{N}\right)$, where $\ket{0(N)}$ are Fock states. In this case entanglement can be calculated directly using the logarithmic negativity (see Appendix \ref{App:RhoCavity}), which shows this result is not unique to our chosen witness. In fact an identical expression has also been found in the linearised regime for Gaussian initial states \cite{miao2020quantum} (in the constant coupling scenario), where it was conjectured that the bound is a generic consequence of the master equation (\ref{maintextmastereq}). Thus the `size' of the superposition created does not matter as both the entanglement and decoherence rates are enhanced in the same way. Note, we have not considered any decoherence on the optical modes. That is, ``shifting'' the entanglement to an apparently protected degree of freedom does not help, as the mechanism used to transfer the correlations also provides a channel for the decoherence to act.

In practice, however, other noise timescales will mean it is still highly desirable to reduce the entanglement/verification time. Thus a large photon number, or more generally photon number variance, is advantageous. On the other hand, the cavities themselves, along with their relative positioning, limits the extent to which the mechanical elements can be driven. These restriction will be discussed in section \ref{Sec:Limitations}.

The analysis for the optimal witness time is straightforward, and we leave the details for Appendix \ref{App:witness_decoherence}. Writing $\kappa^{\rm{dec}}(z(t_q))=\Gamma_{\rm{dec}} z(t_q)$, where from (\ref{main:kappadec}) $\Gamma_{\rm{dec}}=\frac{ 2\Upsilon x_{0,m}^2}{\hbar^2\omega_m \gamma_g } $,  the minimum is found to occur at,
\begin{equation}
D_{\min}(t_q)\approx\frac{1}{\sqrt{2}N_p}\left(1-\frac{3}{2\sqrt{2}}\Gamma_{\rm{dec}}\right), \qquad \Gamma_{\rm{dec}} \ll 1.
\end{equation}
Similarly, the condition for the corresponding witness value to be at least $a$ times the theoretical minimum ($\kappa=0$) is given by,
\begin{equation}
\Gamma_{\rm{dec}} \leq \frac{\sqrt{2}}{5}\left(2-\sqrt{5a -1} +\hdots\right).
\end{equation}
Choosing $a=1/2$ then requires $\Gamma_{\rm{dec}} \lesssim 0.219$, and so $\gamma_R k_BT\lesssim  0.11 \hbar G m/d ^3$. As before, increasing the noise shifts the minimum earlier in time until above $\Gamma_{\rm{dec}} = 1$ when no entanglement can be verified.

For the cavities listed in Table \ref{tab:parameter:values} satisfying the bound (\ref{decoherence_bound}) requires a quality factor $Q\sim 10^{23}$. This is some $15$ orders of magnitude higher than what has currently been achieved (see for example \cite{Matsumoto2019}), and poses a severe obstacle to witnessing gravitational entanglement.

%--------------------------------------------------------------
\subsubsection{Measurement Window}
Given the close connection between $\kappa^{\rm{dec}}(t_q)$ and $D(t_q)$, one may also suspect that $\kappa^{\rm{dec}}$ varies slowly around the measurement time $t_q$. Indeed, we find
\begin{align}
\kappa_{\mu}^{\rm{dec}}(t_q+\zeta\bar{\tau}_m) & \approx  \kappa^{\rm{dec}}(t_q) \\
\nonumber  & \quad +\frac{8\Upsilon x_{0,\mu}^2 }{\hbar^2}\frac{ 4 (n-1)^2(q \pi)^3 n^4 \gamma_g^2 \epsilon^2 \zeta}{(2n-1)^2\omega_m},
\end{align}
where the dependence on the constant (i.e. $k_0$) component enters at order $\gamma_g^2 \zeta^2$. This means the measurement window is dominated by the initial temperature of the state, even if $\bar{n}$ is very small. We can then substitute
\begin{equation}
\kappa(t_q +\zeta\bar{\tau}_m)\approx \Gamma_{\rm{dec}} D(t_q)+2k_0^2 \pi^2(\bar{n}+1/2) \zeta^2,
\end{equation}
into (\ref{kappabound}), where the second term arises from thermal contribution (\ref{kappawindow}), and we have assumed $\kappa^{\rm{th}}(t_q)$ is small (i.e. $\bar{n} \ll \bar{n}_{\max}$). Expanding to second order leads to,
\begin{equation}\label{window_decoherence_thermal}
\delta t \approx \sqrt{\frac{D(t_q)(1-N_pD(t_q)-2\Gamma_{\rm{dec}})}{k_0^2 \omega_m^2(\bar{n}+1/2)}},
\end{equation}
which generalises (\ref{windowthermal}).

%--------------------------------------------------------------
\section{The Casimir effect and Stray Charges}\label{Sec:Casimir}
In order to maximise the gravitational interaction, the mean separation between mechanical elements should be kept as small as possible. This gives the best chance of satisfying (\ref{decoherence_bound}). On the other hand, at small distance scales the Casimir force~\cite{casimir1948influence} becomes significant and can readily dominate over gravity. A direct comparison can be made via the dimensionless coupling parameter $\gamma_{C}\coloneqq\frac{1}{2}\partial_dF_c/m\omega_m^2$, which should be small compared to $\gamma_g$. In general, the Casimir interaction will depend strongly on both the geometry and material properties of the mechanical components. However when the masses are assumed to be spheres, a straightforward estimate can be made in the zero temperature regime, provided the separation is much larger than the radii, $d\gg R$. Using equation (21) of \cite{Rodriguez-Lopez:2011}, one finds that, 
\begin{equation}
\gamma_C=\frac{161\hbar c R^6}{\pi m \omega_m^2 d^9}.
\end{equation}
The requirement that $\gamma_g \ll \gamma_{C}$ immediately implies $m \gg   \sqrt{\frac{161}{\pi}} \left(\frac{R}{d}\right)^3 m_p$,
where $m_p=\sqrt{\hbar c/G} \approx 2.176 \times 10^{-8}$kg is the Planck mass. This suggests that the masses in general should be (relatively) large, otherwise the separation must increase to compensate. Indeed, for the parameters in Table \ref{tab:parameter:values}, this leads to $\gamma_C \sim 1\times 10^{-17}$ and $\gamma_g\sim 2\times 10^{-18} $, and so Casimir mediated entanglement would dominate over gravitational effects. On the other hand, a large mass will also be harder to cool which can lead to shortening of the available measurement window, c.f. (\ref{window_decoherence_thermal}). 

To get around this problem, we can adopt an approach frequently used in small distance tests of gravity. This involves introducing a suitable shielding material, for example gold \cite{chiaverini2003new}, between a driven source and detector. If the material is rigid enough, then the only oscillating signal is that from the gravitational interaction, which can be identified over the stronger (constant) Casimir/electrostatic background. A similar screening effect has been proposed for entanglement generation \cite{Wan2017,vandeKamp2020}, where the transmission of quantum correlations are effectively suppressed because any induced motion in the shield, including its position variance, will be small. In this sense, we can consider the shield as restricting non-gravitational forces to only a classical channel.

It is worth pointing out that the witness (\ref{maintextWit1}) itself also provides some limited protection against false positives. We note that our previous results can be immediately generalised to $\gamma_g \rightarrow \gamma_{\rm{tot}}$, where $\gamma_{\rm{tot}}=\frac{1}{2}\sum_i \partial_dF_i/m \omega_m^2$ now accounts for all $\hat{x}_1\hat{x}_2$ couplings in the system. For example, electrostatic interactions can be included via the Coulomb force, $F_Q$, leading directly to the parameter $\gamma_Q=\frac{q_1 q_2}{4\pi \varepsilon_0 m d^3 \omega_m^2}$, where $q_1$ and $q_2$ are charges on the mechanical oscillators, and $\varepsilon_0$ is the vacuum permittivity. Thus, in the absence of significant shielding, we can write $\gamma_{\rm{tot}}=\gamma_g+\gamma_{C} +\gamma_Q$. Now as $D(t_q) \propto \gamma_{\rm{tot}}$, if one of the additional interaction parameters is comparable to $\gamma_{g}$, then $D(t_q)$ will double, leading to a higher than expected entanglement rate. However, as can be seen from figure \ref{fig:witness}, beyond a certain time the witness is positive. We can estimate this from (\ref{kappabound}) along with the relevant $\kappa(t_q)$. For example, in the case of constant optomechanical coupling (\ref{kappathermalz}) (and neglecting thermal decoherence), a second order expansion in $z$ leads to the condition,
\begin{equation}\label{maxtime}
z(t_q) \lesssim \pm \frac{1}{1 + 2N_p +\Gamma_0^{\rm{th}}}.
\end{equation}
This tells us that above $|D(t_q)| \sim 1/(N_p +\Gamma_0^{\rm{th}}/2)$ the witness is positive, and so from (\ref{Dconstopt}), if $\gamma_{\rm{tot}}$ is larger than expected by $\sim \sqrt{2}$ an experiment aiming to measure at the optimal time will fail to detect entanglement, even if it is present. For the parameters in table \ref{tab:parameter:values} this would already happen for only $0.006$ stray electrons on each mass (ignoring the Casimir interaction). Of course, as $z(t)=\sin{(D(t)/2)}$, there are further periodic solutions, for example when $D(t_q)$ is shifted to $\tilde{D}(t_q)=2\pi\pm D(t_q)$. However, in this case $\gamma_{\rm{tot}}$ would need to be larger by a factor $\sim N_p$. It then also becomes much harder to satisfy $\kappa < z$ (from (\ref{kappabound})), as $\kappa^{\rm{th}}(t_q)$ grows quadratically in $\gamma_{\rm{tot}}$. Thus, in principle, it should be possible to engineer an experiment such that if entanglement is witnessed, we can be reasonably confident it did not emerge from an undesired interaction.

%--------------------------------------------------------------
\section{Limitations due to the mechanical motion}\label{Sec:Limitations}

In equation (\ref{eq:systemHamiltonian}), we started from a Hamiltonian description of the light-matter coupling based on a perturbative treatment of the motion of the mechanics \cite{law_interaction_1995, romero2011optically, serafini2017quantum}. This means our results can only be seen as reliable if the motion of the mechanical elements is small. For example, in Fabry--P\'{e}rot cavities the displacement of the end mirror must be much smaller than the cavity length, while in levitated systems the relevant lengthscale is set by the wavelength of the light modes~\cite{Millen_2020}. Similarly, the gravitational interaction between the mirrors was described by a second order expansion of the Newtonian potential, which requires their displacements to be much smaller than their mean separation. In principle, higher order treatments can be possible (see, for example \cite{bruschi2020time}), however the dynamics will always be limited by the physical dimensions of the experiment.

As the gravitational interaction is weak, by far the strongest effects on the mechanics comes from the radiation pressure of the cavity fields. This shifts not only the equilibrium position, but also increases the oscillation amplitudes of the mechanical elements. The average radiation pressure force, however, can be cancelled using an appropriately tailored external potential (which enters through the $C_{\mu}$ coefficients, (\ref{eq:F:coeffs})). Thus the mean position can be set to zero (see (\ref{App:positions})), and all that remains are the quantum fluctuations of the mechanical motion. The corresponding variances for the center of mass mode and the stretch mode are (see appendix \ref{sec:displ})
\begin{align}
\nonumber (\Delta \hat{x}_c)^2 & = (\Delta \hat{x}_{\omega_c})^2 \\
\nonumber &  \quad + 4 x_{0,c}^2 \Re[ e^{-i\omega_s t} K_{\hat{N}_c}]^2 \left(\Delta (\hat{N}_2-\hat{N}_1)\right)^2, \\ (\Delta   \hat{x}_s)^2 &= (\Delta \hat{x}_{\omega_s})^2 \\
\nonumber  & \quad  + 4 x_{0,s}^2 \Re [ e^{-i\omega_s t} K_{\hat{N}_s}]^2  \left(\Delta (\hat{N}_2+\hat{N}_1)\right)^2 ,
\end{align}
where
\begin{equation}
\begin{aligned}
\hat{x}_{\omega_\mu}(t)&=x_{0,\mu}\left(e^{i\omega_\mu t}\bd_\mu + e^{-i\omega_\mu t}\hat{b}_\mu\right)\,,
\end{aligned}
\end{equation}
is the co-rotated position operator of the mode $k$ and for a thermal state $(\Delta \hat{x}_{\omega_\mu})^2 = (2\bar{n}_\mu+1)x_{0,\mu}^2$.

Imposing the condition $\Delta \hat{x}_c, \Delta \hat{x}_s \ll d$ (and similarly for the cavity length $L$) implies a limit on the thermal occupation and the photon number variance. The former has to be minimized in any case to detect entanglement generation, however it is generally advantageous to maximise the latter and so its restriction needs to be accounted for when choosing system parameters. 

For the modulated optomechanical coupling at frequency $\Omega_n = (1-1/n)(\omega_c+\omega_s)/2$, ($\phi_1=\pi/2$), we find at zeroth order in $\gamma_g$,
\begin{widetext}
	\begin{align}
		\Re [e^{-i\omega_s t} K_{\hat{N}_s}(t)]^2  &\approx \frac{\left( k_0(2n-1)(1-\cos(\omega_m t)) + \epsilon n (n-1) \sin(\omega_m t) - \epsilon n^2 \sin((1-1/n)\omega_m t)\right)^2}{2(2n-1)^2} \le  \frac{\left( 2k_0 + \epsilon n  \right)^2 }{2} \,.
	\end{align}
\end{widetext} 

This implies the conditions ${\sqrt{2}}x_0(2k_0 + \epsilon n)\Delta\hat{N}_{1,2} \ll d$, or formulated as a constraint on the photon number standard deviations: $\Delta\hat{N}_{1,2} \ll d/(\sqrt{2}x_0(2k_0 + \epsilon n))$. For the values in Table \ref{tab:parameter:values}, we find $\Delta\hat{N}_{1,2}\ll 3\times 10^6$. Since for coherent states, $\Delta\hat{N}^2=\langle \hat{N} \rangle$, this condition is fulfilled.

%-------------------------------------------
\section{Comparison to Semi-Classical Analysis}\label{Sec:semiclassical}
%---------------------------------
For ancilla based entanglement tests, an alternative approach for analysing the entanglement rate is to work in a semi-classical approximation. This has the advantage that the full, coupled, dynamics need not be solved (in fact, an explicit Hamiltonian is not even required). In the case of optomechanical systems, this turns out to be remarkably accurate when only second order perturbations are included.

We start by constructing a model for two massive particles, where each exists in superpositions of approximately classical trajectories $x_{1,n}(t)$ and $x_{2,n}(t)$. These paths are labelled by sets of orthonormal quantum states, $\ket{u_{n}}$ and $\ket{v_{n}}$, respectively, encoded on two additional (external or internal) degrees of freedom. We can write the total state of each system as, 
	\begin{align}
	\ket{\psi_1}&=\sum a_{n} \ket{u_{n}}\ket{\Phi_{1,n}(t)},\,\,\mathrm{and}\nonumber\\
	\ket{\psi_2}&=\sum b_{n} \ket{v_{n}}\ket{\Phi_{2,n}(t)},
	\end{align}
where $\ket{\Phi_{i,n}}$ are the mechanical states corresponding to the trajectories $x_{i,n}(t)=\avg{\hat x_i}_{\ket{\Phi_{i,n}(t)}}$. The gravitational interaction between the two particles can be modelled by assuming the potential energy difference between each localised superposition gives rise to an accumulating phase,
\begin{equation}\label{semiclassicalphases}
 \phi_{nm}(t)= \frac{Gm^2}{\hbar} \int^t_0 \frac{dt'}{x_{2,m}(t') - x_{1,n}(t')}\,,
\end{equation}
and so after some time, the total state of the combined system is, 
\begin{equation}
\ket{\psi(t)} = \sum_{n,m} a_{n} b_{m} e^{i\phi_{nm}} \ket{u_{n}}\ket{v_{m}}\ket{\Phi_{1,n}(t)}\ket{\Phi_{2,m}(t)}.
\end{equation}
	
Note that entanglement only enters at the second order expansion (and above) of (\ref{semiclassicalphases}). We now make the assumption that after a period $t_q$ the mechanical and label states disentangle, such that the reduced state of the total ancilla system is pure,
\begin{equation} \ket{\psi_\mathrm{red}(t_q)} = \sum_{n,m} a_{n} b_{m} e^{i\phi_{nm}} \ket{u_{n}}\ket{v_{m}}.
\end{equation}  
For low dimensional systems, the entanglement can be calculated directly from the above. However, the phases $\phi_{nm}$ can also be viewed as the eigenvalues of a suitable operator with eigenstates $\ket{u_{n}}\ket{v_{m}}$. The former is dependent on the difference between classical trajectories, which suggests a natural construction is to first define two local operators, 
\begin{equation}
\begin{aligned}\label{Oconstruction}
\hat{O}_1 =\sum_n u_n\ketbra{u_n}{u_n}, \quad \text{and}\quad 	\hat{O}_2 =\sum_n v_n\ketbra{v_n}{v_n}.
\end{aligned}
\end{equation}
If the paths can be written as functions of the label eigenvalues $u_n$, and $v_m$, then the desired operator can be found from $\Delta x(t,\hat{O}_1,\hat{O}_2)$, where $\Delta x(t,u_n,v_m):= x_{2,m}(t) - x_{1,n}(t)$. Thus the time evolution of label states (at the decoupling time) is of the form,
\begin{equation}
\hat{U}= \hat{U}_1 \hat{U}_2 e^{i \frac{Gm^2}{\hbar} \int^t_0 \Delta x(t,\hat{O}_1,\hat{O}_2)^{-1} dt'} .
\end{equation}
By expanding $\Delta x(t,\hat{O}_1,\hat{O}_2)^{-1}$ to lowest order such that the time evolution operator can be expressed as
\begin{equation}
\hat{U}= \hat{\tilde{U}}_1 \hat{\tilde{U}}_2 \hat{U}_{12} = \hat{\tilde{U}}_1 \hat{\tilde{U}}_2 e^{-i  D \hat{O}_1 \hat{O}_2  } ,
\end{equation}
we can make direct use of (\ref{SLapprox}) and the results of Appendix \ref{App:LinearEntropy}. 

\emph{Example 1: Interferometry.} One of the simplest examples has been presented in \cite{Bose2017:spin}, where each particle is instantaneously driven into a cat-state superposition with separation $\iota$, and then returned to its original state after time $t$. We can define label states $\ket{u_\pm} =\ket{v_\pm} \equiv \ket{\pm\frac{\iota}{2}}$, then the interaction operator is given by, 
\begin{equation}
U_{12}(t)=e^{-i\frac{2 Gm^2}{ \hbar d^3}\hat{O}_1\hat{O}_2 t},
\end{equation}
where $\hat{O}_1$ and $\hat{O}_2$ are constructed from (\ref{Oconstruction}).
For a two level system, the linear entropy is a function of only $\Delta \hat{O}_1 \Delta \hat{O}_2 = \frac{\iota^2}{4}$ and so from (\ref{SL_twolevel}) the condition for maximal entanglement is \cite{Bose2017:spin},  
\begin{equation}
\frac{ Gm^2}{ 2\hbar d}\left(\frac{\iota}{d}\right)^2 t = \pi.
\end{equation}
A characteristic time scale for more general initial states (at least to reach small levels of entanglement) can easily be found from the analogue of (\ref{SL}) or, to higher order (\ref{LinEnt}).

\emph{Example 2: Optomechanical systems.} In this case the difference between classical trajectories is given by $\Delta x(t,u_n,v_m) = d + \sqrt{2} x_s(t,u_n,v_m)$, which for thermal mechanical states can be readily found from (see Appendix \ref{sec:displ}),
\begin{align}
	\nonumber 	x_s(t,u_n,v_m)  &=  2 x_{0,s}\Big(\Re [ e^{-i\omega_s t} C_s] - \Re [e^{-i\omega_s t}K_{\hat{N}_s}] \\
  & \quad \times \left(\avg{\hat{N}_1}_{\ket{u_n}}+\avg{\hat{N}_2}_{\ket{v_m}}\right)\Big).
\end{align}
As this depends on the photon number of the ancilla (cavity) state, we set $\hat{O}_i=\hat{N}_i$. A second order expansion of $\Delta x(t,\hat{O}_1,\hat{O}_2)^{-1}$ then leads to the time evolution operator,
\begin{equation}
U= \tilde{U}_1 \tilde{U}_2 e^{-i \frac{16Gm^2}{\hbar d^3} x_{0,s}^2 \int_0^t \Re [e^{-i\omega_s t'}K_{\hat{N}_s}(t')]^2 dt' \hat{N}_1 \hat{N}_2    },
\end{equation}
where we can identify,
\begin{align}
  D(t) &= \frac{16Gm^2}{\hbar d^3}   x_{0,s}^2 \int_0^t \Re [e^{-i\omega_s t'}K_{\hat{N}_s}(t')]^2 dt'  .
\end{align}
For both modulated and unmodulated optomechanical couplings we find that $D(t_q)$ evaluates to (\ref{eq:Dconst}) and (\ref{eq:Dfrac}) to first order in $\gamma_g$. The entanglement rate estimates of section \ref{Sec:EntanglementTime} then follow immediately. Note however, that at higher order in $\gamma_g$, deviations between the fully quantum and semi-classical treatments begin to appear.

%-------------------------------------------
\section{Quantum Metrology}\label{Sec:Metrology}
We can gain further intuition for the observed initial state dependence by making use of techniques from quantum metrology. In this framework, we consider the two cavities as a sensor-source pair. Then a prerequisite for any experiment is that its sensitivity to gravity is high enough to resolve quantum features in the source. In our case, fluctuations of the photon number.

Formalising this perspective is complicated by the fact that there is no parameter corresponding to the superposition size in the evolution operator (\ref{Ugeneral}), even within the semiclassical model discussed in section \ref{Sec:semiclassical}. 
This prevents a direct application of the usual methods from metrology. One solution is to first characterise the measurement precision for the case that $\ket{\psi_{\rm{source}}(0)}$ is an eigenstate of one of the quantum degrees of freedom, i.e. $\hat{N}_{\rm{source}}$. Then we can effectively replace the operator by its eigenvalue, $\lambda_N$, in $\hat{U}$ and treat this as a parameter for estimation. The quantum Fisher information (QFI) of the evolved state then sets the ultimate lower limit on the variance of an unbiased estimator $\tilde{\lambda}_N$ via the Cram\'er-Rao bound, 
\begin{equation}\label{QFI}
(\Delta \tilde{\lambda}_N)^2 \geq \frac{1}{\mathcal{N}{\rm{QFI}}}.
\end{equation}
where $\mathcal{N}$ is the number of measurements. We then argue that resolving quantum fluctuations in the source photon number is only possible when $\Delta \hat{N}_{\rm{source}} \geq \Delta \hat{\lambda}_N$.

The QFI is most easily evaluated when the cavity field states are pure. While this will strictly never be true, it is comparable to the approximations used in section \ref{Sec:EntanglementTime} where we assumed $K_{\hat{N}_\mu}(t_q)\approx0$. In this case it takes the simple form,
\begin{equation}
{\rm{QFI}}=4(\avg{\psi'_{\lambda_N}|\psi'_{\lambda_N}}-|\avg{\psi'_{\lambda_N}|\psi_{\lambda_N}}|^2),
\end{equation}
where $\ket{\psi_{\lambda_N}}=\hat{U}\ket{\psi(0)}$ is the field state of the detector at $t_q$, and $\ket{\psi'_{\lambda_N}}$ is its derivative with respect to the parameter $\lambda_N$. Denoting $\hat{N}_2=\hat{N}_{\rm{detector}}$ and $\hat{N}_1\rightarrow \lambda_N$ in (\ref{Ugeneral}), a straightforward calculation leads to,
\begin{equation}
{\rm{QFI}}(t_q)=4 D(t_q)^2 (\Delta \hat{N}_{\rm{detector}})^2,
\end{equation}
(neglecting $K_{\hat{N}_\mu}$ contributions). This is a direct consequence of the approximation, $\hat{U}_{12}(t_q)=\hat{U}_1\hat{U}_2e^{iD(t_q) \hat{N}_1\hat{N}_2}$. It then follows from (\ref{QFI}), and the condition $\Delta \hat{N}_{\rm{source}} \geq \Delta \hat{\lambda}_N$, that,
\begin{equation}\label{Dmetrology}
|D(t_q)| \geq \frac{1}{2 \Delta \hat{N}_{\rm{source}} \Delta \hat{N}_{\rm{detector}}}\,.
\end{equation}
Thus we see that metrology predicts a timescale that is broadly in line with that needed for entanglement generation (\ref{SL}). This should come as no surprise as the semiclassical model above tells us that entanglement is only possible when (at least some of) the phases $\phi_{nm}$ are different. On the other hand, to distinguish between them, we should be able to resolve differences in the corresponding trajectories, which in turn depend on fluctuations of the photon number. 

To make this link more apparent, we can instead derive bounds explicitly for the mirror position. The result turns out to be essentially the same whether we use expectation values or eigenvalues to identify a suitable estimator.
Note, while our focus has been on entanglement generation, we could equally interpret the sensitivity requirements as those needed to test whether gravity respects the superposition principle\footnote{in fact we can ignore the quantum back-action entirely from the analysis.}. The latter can be traced back to proposals by Feynman \cite{cecile2011role}, but has also received renewed attention in recent years \cite{Anastopoulos2015,Carlesso2019,Anastopoulos_2020}.

To begin, we consider measuring the classical position $x_{\rm{source}}$ of a particle (of mass $m$) using a single optomechanical system. In the rotating frame, the detector Hamiltonian takes the form (\ref{eq:modulatedH}), ($\mu \rightarrow \rm{detector}$) where by analogy with (\ref{2particleHamiltonian_Newtonian}), the displacement due to the gravitational interaction is given by
\begin{equation}\label{Ddetector}
\mathcal{D}_{\rm{detector}}(t) = \frac{\gamma_g }{ 2 x_{0,\rm{detector}}}(d-2 x_{\rm{source}}).
\end{equation}

It has been shown in \cite{Qvarfort2021} and \cite{Schneiter2020opt} that the QFI of the local (pure) field state, with respect to the parameter $x$ is given by 
\begin{equation}
{\rm{QFI}}= 4 W_{x_1}^2 (\Delta \hat{N}_{\rm{detector}})^2,
\end{equation}
where 
\begin{equation}
W_{x_1} = -\partial_{x_1} F_{\hat{N}_2}  - 2F_{\hat{N}_2 \hat{B}_{2,-}} \partial_{x_1} F_{\hat{B}_{2,+}} ,
\end{equation}
(for notational convenience we abbreviate source and detector labels to $1$ and $2$ respectively). The $F$-coefficients can be determined by substituting (\ref{Ddetector}) into formally identical equations to (\ref{eq:F:coeffs}). Note, in our case the position (as an estimation parameter) has a time dependence, $t$, which is independent of the actual measurement time, $t_q$. Indeed, if $x_{\rm{source}}(t) \equiv \avg{\hat{x}_1(t)}$, then for a thermal state (see appendix \ref{App:positions})
\begin{equation}
x_{\rm{source}}(t)= 2 x_{0,1}(\Re [e^{-i\omega_m t} C_1]+ \Re[e^{-i\omega_m t}K_{\hat{N}_1} ]) \avg{\hat{N}_1},
\end{equation}
which can also be zero. Thus, it is convenient to adopt the new parameter $N_1 \equiv\avg{\hat{N}_1}$ and write $W_{x_1}= (\partial N_1/\partial x_1)W_{N_1}$. The calculation can be simplified further if we ignore any displacement of the source due to the detector (i.e. setting $\mathcal{D}_{\rm{1}}(t))=0$), which means $C_1(t)=0$. Then evaluating $W_{N_1}$ when both source and detector systems are identical, we find to first order in $\gamma_g$,  
\begin{align}
\nonumber W_{N_1}(t_q) &\approx  4 \pi n q \gamma_g \left(3 k_0^2 + \epsilon^2\frac{n^2( 1 + (2n-1)^2 )}{2(2n-1)^2}\right)\\
& \approx   D(t_q),
\end{align}
which is not a function of the time $t$. Thus the Cram\'er-Rao bound for an unbiased estimate of $x_{\rm{source}}(t)$ using measurements at decoupling time $t_q=2\pi n q/\omega_m$ is given by,
\begin{equation}\label{CRposition}
(\Delta \tilde{x}_{\rm{source}}(t) )^2 \gtrsim \frac{4 x_0^2 \Re[ e^{-i\omega_m t} K_{\hat{N}_{\rm{source}}}(t)]^2}{4D^2(t_q) (\Delta \hat{N}_{\rm{detector}})^2}.
\end{equation}

We now argue that superpositions of the gravitational field can only be resolved when the detector precision, (\ref{CRposition}), is smaller than the variance of the position operator $\hat{x}_{\rm{source}}$,
\begin{equation}
\begin{aligned}
(\Delta \hat{x}_{\rm{source}}(t))^2 &=4 x_{0}^2 {\Re[e^{-i\omega_m t} K_{\hat{N}_{\rm{source}}}(t)]^2} (\Delta \hat{N}_{\rm{source}})^2 \\
 &\quad +(\Delta \hat{x}_{\omega}(t))^2 ,
\end{aligned}
\end{equation}
where,
\begin{equation}
    \hat{x}_{\omega}(t)=\hat{x}_{\rm{source}}(0)\cos \omega_m t + \frac{1}{m \omega_m}\hat{p}_{\rm{source}}(0) \sin \omega_m t.
\end{equation}
If we are interested in the entanglement between cavity fields, then the important correlations are those arising between the respective number operators. We can therefore ignore the second term above, and the condition $\Delta \hat{x}_{\rm{source}}(t) \geq \Delta \tilde{x}_{\rm{source}}(t)$ immediately recovers (\ref{Dmetrology}).

%-----------------------------
\section{Conclusions}
The application of the LOCC framework to tests of quantum gravity has gained significant traction over the last few years. In particular, because direct access to the state of the gravitational field – for example, by probing the existence of gravitons – is almost certainly unachievable. Instead, it is hoped that correlations arising from the collective interactions of macroscopic-scale quantised masses may provide indirect evidence for the nature of gravity. 

Such tests are exceptionally difficult, requiring highly non-classical states or very long integration times. This poses a challenge for traditional optomechanical implementations, where generating large cat-states requires superpositions of very different photon numbers. Instead one must typically resort to a combination of low mechanical frequencies and high coupling strengths. 

In contrast we have shown that the entanglement rate can be significantly increased if the optomechanical coupling can instead be modulated close to resonance \footnote{
In principle, it is also possible to use modulations of the mechanical frequency or the interaction strength to enhance the growth of entanglement. We leave this to future work.}. For $k_0=\epsilon$, this advantage scales as roughly $\frac{1}{12}\left(\frac{\omega_m^2 d^3}{G m \pi N_p }\right)^{\frac{2}{3}}$ -- i.e. the benefit is greatest precisely in the most accessible parameter regime. This can be traced back to the position response of the differential mode, which to a good approximation governs the coefficient $D(t)$ which controls the non-separability of the cavity states (see first appearance in Eq.\eqref{Ugeneral}), and not the initial cavity state. The latter can of course be leveraged for further improvements. For example, (\ref{SL}) can also be readily applied to squeezed coherent states \cite{walls1983squeezed}, where for squeezing amplitude $r$ and coherent state parameter $\alpha$, the photon number and its variance are given by $N_p = |\alpha|^2 \, e^{2r} + \sinh^2(r)$ and $(\Delta \hat N_j)^2 = |\alpha|^2 e^{4r} + \frac{1}{2}\sinh^2 (2r)$, respectively (for appropriately chosen phases) \cite{Qvarfort2021}. For squeezing of around $10$ dB \cite{ast2013high}, corresponding to $r\sim 1.73$, the requirement on $D(\tau_e)$ would be decreased by an additional factor of $30$.

In the absence of noise, modulations would therefore suggest that entanglement timescales comparable to proposals based on mechanical cat-states, (e.g. \cite{Bose2017:spin}) may be realistic. However, even with these improvements, significant obstacles remain. First, the measurement window for witnessing entanglement is extremely sensitive to the initial temperature of the mechanical modes. Similar behaviour is well known for the single photon interference visibility width in a Michelson style optomechanical interferometer \cite{Marshall2003,Bernad2006}, which scales roughly as $\sqrt{1/\bar{n}}$. In our case, the additional photon number dependence in (\ref{windowthermalopt}) is particularly problematic if entanglement is to be achieved on short timescales. To some extent, this can be alleviated through purely oscillatory couplings (i.e. $k_0=0$), which reduces the dependence to $\sim 1/(N_p\bar{n})^{1/4}$. However, in practice this still means that the system must be cooled far below the limits set by (\ref{nmax_mod}), even for state of the art detector bandwidths.

Secondly, in a two cavity system, the relative dynamics must be matched to extremely high precision. This guarantees that both mirrors simultaneously disentangle from the optical modes (at least approximately), maximising the measurable entanglement. Provided both cavities are initialised at the same time, the main determiner is whether the mechanical frequencies are equal, see section \ref{Sec:frequency_mismatch}. Even for instantaneous measurements, the agreement should still be better than one part in $2\pi k_0 \sqrt{N_p (\bar{n}+\frac{1}{2})}$, which is a significant challenge. 

The most serious issue, however, comes from environmental decoherence. For both modulated and unmodulated couplings, we find that the relaxation rate must be bounded by $\gamma_R \leq \frac{\hbar}{2 k_B T} \frac{Gm}{d^3}$. This agrees with results previously reported in other scenarios, for example in two free oscillators \cite{Kafri_2014}, linearised optomechanics \cite{miao2020quantum} and levitated nano systems \cite{Rijavec_2021}, which supports the expectation that this is inherently a property of the noise model (\ref{maintextmastereq}). That is, the constraint is independent of both the local dynamics and initial conditions -- it cannot be improved upon with novel quantum control. To put this in context, even using the highest density materials ($2.2\times 10^4$kg/m$^3$) and milliKelvin temperatures, the relaxation time needs to be longer than $\gamma_R^{-1} =Q/\omega_m \approx 10^{14}$s. The requirement on the $Q$ factor can be reduced with low frequency mirrors, however cooling then becomes more demanding. 

There are two caveats here. In practice, equation (\ref{maintextmastereq}) is likely to be too simplistic a description, and there is evidence that, at least in some optomechanical systems, the environment turns out to be non-Markovian \cite{Groblacher2015}. This is likely to have a further negative impact entanglement generation \cite{Ludwig2010}. On the other hand, there are also suggestions that the steady state limit of the full quantum Brownian motion model is more protected against loss of coherence \cite{Datta2021}. Whether this holds in the non-linear regime and for sufficiently short timescales -- particularly in macroscopic systems where the discounted terms will typically be much smaller than those in (\ref{maintextmastereq}) -- remains to be seen. Alternatively, one may also look to weaker quantum correlations. However, in this case the motivation is less clear, as one no longer has the guarantee of monotonicity under LOCC (e.g. Discord \cite{dakic2010necessary}).

%-----------------------------
\section*{Acknowledgements}

The authors would like to thank Sofia Qvarfort and Stefan Scheel for helpful remarks and discussions. A.~D.~K.~P. was partially supported by the European Social Fund (ESF) and the Ministry of
Education, Science and Culture of Mecklenburg-Western Pomerania (Germany) within the project NEISS under grant no ESF/14-BM-A55-0006/19. D.R. acknowledges funding by the Federal Ministry of Education and Research of Germany in the project “Open6GHub” (grant number: 16KISK016) and support through the Deutsche Forschungsgemeinschaft (DFG, German Research Foundation) under Germany’s Excellence Strategy – EXC-2123 QuantumFrontiers – 390837967, the Research Training Group 1620 “Models of Gravity” and the TerraQ initiative from the Deutsche Forschungsgemeinschaft (DFG, German Research Foundation) – Project-ID 434617780 – SFB 1464.

\bibliographystyle{quantum}
\bibliography{entanglement}

%--------------------------------- Appendices ---------------------------------
\onecolumngrid
\appendix
  
%----------------------------------------
\section{Linear Entropy for restricted interaction}\label{App:LinearEntropy}

In order to estimate an upper limit to the entanglement rate, we will find it convenient to work with the linear entropy, $S_L= 1-\tr[\rho_1^2]$, where $\rho_1=\tr_2 \rho$ is the reduced state of a bipartite system (in our case, the joint cavity state for the two field modes). A characteristic entanglement timescale can then be defined as the time, $\tau_e$, for $S_L$ to reach some fixed value, typically chosen based on the achievable signal to noise ratio of the measurement. Here we will show that for pure separable states, $\rho_0=\rho_1\otimes \rho_2$, the linear entropy after the evolution $\hat{U}=\hat{U}_1\hat{U}_2e^{iD\hat{O}_1 \hat{O}_2}$ is given by,
\begin{equation}\label{LinEnt}
S_L= -\sum_{k=1}^{\infty}\frac{(-1)^k D^{2k} }{(2k)!} \sum_{s=0}^{2k}\colvec{2}{2k}{s}(-1)^{s} \avg{\hat{O}_1^{2k-s}}\avg{\hat{O}_1^s}\sum_{r=0}^{2k}\colvec{2}{2k}{r}(-1)^{r} \avg{\hat{O}_2^{2k-r}}\avg{\hat{O}_2^r},
\end{equation}
where $\hat{O}_j$ ($\hat{U}_j$) are arbitrary Hermitian operators (unitaries) acting on the local systems, and the expectation values are taken with respect to the initial states $\rho_1$ and $\rho_2$.

The first step is to write $\rho_j=\ketbra{\psi_j}{\psi_j}$, then the purity of the reduced state of subsystem one is given by, 
\begin{equation}
\tr[\rho_1^2]= \sum_{r,s} |\avg{\psi_1,\psi_2|\hat{U}^{\dagger}|w_r} \avg{w_s|\hat{U}|\psi_1,\psi_2}|^2,
\end{equation}
where $\{\ket{w_r}\}$ is any complete basis of $\mathcal{H}_2$. We now expand $\ket{\psi_1}=\sum_p a_p\ket{u_p}$ in terms of the eigenstates, $\ket{u_p}$, of the operator $\hat{O}_1$ (note the spectra can be continuous). For the form of $\hat{U}$ above, the terms on the right hand side can be written as,
\begin{equation}
\bra{w_s} \hat{U} \ket{\psi_1,\psi_2} = \sum_{p} a_p \hat{U}_1 \bra{w_s}\hat{U}_2 e^{iD\hat{O}_1 \hat{O}_2} \ket{\psi_2} \ket{u_p},
\end{equation}
and so the purity is given by,
\begin{equation}
\tr[\rho_1^2]= \sum_{r,s} \left|\sum_p |a_p|^2\avg{\psi_p|U_2^{\dagger}|w_r} \avg{w_s|U_2 |\psi_p}\right|^2,
\end{equation}
where we have defined $\ket{\psi_p}=e^{iDu_p\hat{O}_2 }\ket{\psi_2} \in \mathcal{H}_2$. The factors appearing after the summation are now all c-numbers, and so by expanding the square, we can remove the sums over $r$ and $s$ to leave, 
\begin{equation}\label{App:Purity1}
\tr[\rho_1^2]= \sum_{p,q} |a_p|^2 |a_q|^2 |\avg{\psi_p|\psi_q}|^2.
\end{equation}
Now, writing $\ket{\psi_2}=\sum_m b_m\ket{v_m}$ in terms of the eigenstates of $\hat{O}_2$, we have,
\begin{equation}
\begin{aligned}
|\avg{\psi_p|\psi_q}|^2&= \sum_{m,n} |b_m|^2|b_n|^2 e^{iD(v_m-v_n)(u_p-u_q)} \\
&=\sum_{m,n} |b_m|^2|b_n|^2 \cos\left(D(v_m-v_n)(u_p-u_q)\right),
\end{aligned}	
\end{equation}
where simplification in the last line arises because imaginary parts of the exponential must cancel in order for the right hand side to be real. Inserting into (\ref{App:Purity1}), the linear entropy can be expressed as, 
\begin{equation}\label{LinEntropy}
S_L= 1-\sum_{p,q,m,n} |a_p|^2 |a_q|^2 |b_m|^2|b_n|^2 \cos\left(D(v_m-v_n)(u_p-u_q)\right).
\end{equation}
We now note that expectation values of powers of the operator $\hat{O}_1$ are given by, 
\begin{equation}
\avg{\hat{O}_1^r}_{\psi_1} = \sum_n |a_n|^2 u_n^r,
\end{equation}
and similarly for $\hat{O}_2$. Thus, expanding the cosine appearing in (\ref{LinEntropy}) as the power series $\cos(x) =1+\sum_{k=1} \frac{(-1)^k x^{2k}}{(2k)!}$ requires evaluating terms of the form,
\begin{equation}
\begin{aligned}
y_j&=\sum_{p,q} |a_p|^2 |a_q|^2 (u_p-u_q)^j \\
&= \sum_{p,q} |a_p|^2 |a_q|^2 \sum_{s=0}^{j}\colvec{2}{j}{s}u_p^{j-s}(-1)^su_q^s\\
&=\sum_{s=0}^{j}\colvec{2}{j}{s}(-1)^s \avg{\hat{O}_1^{j-s}}\avg{\hat{O}_1^s}.
\end{aligned}
\end{equation}
Similar terms can also be found for the eigenvalues $v_j$, and so finally we have (\ref{LinEnt}),
	\begin{equation*}\label{App:LinEnt}
	S_L= -\sum_{k=1}\frac{(-1)^k D^{2k} }{(2k)!} \sum_{s=0}^{2k}\colvec{2}{2k}{s}(-1)^{s} \avg{\hat{O}_1^{2k-s}}\avg{\hat{O}_1^s}\sum_{r=0}^{2k}\colvec{2}{2k}{r}(-1)^{r} \avg{\hat{O}_2^{2k-r}}\avg{\hat{O}_2^r}.
	\end{equation*}

Ideally, the measurement scheme should be sensitive to small amounts of entanglement, then it is sufficient to consider only the lowest order expansion in $D$,
\begin{equation}\label{App:SLseries}
S_L\approx 2D^2 (\Delta \hat{O}_1)^2 (\Delta \hat{O}_2)^2.
\end{equation}
In this case, the time needed to generate verifiable entanglement can be found by solving the equation
\begin{equation}\label{App:SL}
|D(\tau_e)| \approx \frac{c}{ \Delta \hat{O}_1 \Delta \hat{O}_2},
\end{equation}
where the constant, $c$, depends on the limits imposed by detector noise (we address the separate question on noise in the quantum state in section \ref{sec:verification}). Thus, for a fixed sensitivity, the entanglement rate is enhanced by engineering initial states with large variances for the coupling operators. For some states, the scaling in (\ref{App:SL}) effectively extends to all values of the linear entropy. Setting $\hat{O}_j = \hat{N}_j$, then one can readily confirm that for Fock state superpositions of the form $\ket{\psi}_j=\frac{1}{\sqrt{2}}(\ket{0}_j+\ket{N}_j)$, the linear entropy in (\ref{LinEnt}) is given exactly by, $S_L=\frac{1}{4}(1-\cos DN^2)$\footnote{More generally, for normalised states $\ket{\psi_{1}}=a_1\ket{m}+a_2\ket{n}$ and $\ket{\psi_{2}}=b_1\ket{p}+b_2\ket{q}$ the linear entropy is given by, 
	\begin{equation}\label{SL_twolevel}
	S_L=4 |a_1|^2 |a_2|^2 |b_1|^2 |b_2|^2 (1-\cos D(m-n)(p-q)).
	\end{equation}
So while the reduced state may not always be maximally mixed, the condition for the linear entropy to be maximised is still $D(m-n)(p-q)=\pi$. This also shows that not all couplings lead to a maximally entangled state, for which $S_L=1-\frac{1}{\text{dimension}}=\frac{1}{2}$.}. Therefore (\ref{SL}) is satisfied with $c=\frac{1}{4}\cos^{-1}(1-4S_L)$, for $0\leq S_L \leq\frac{1}{2}$, and in particular the maximum entanglement scales as $D=\pi/N^2$. For fixed photon numbers, $N_p \equiv \avg{\hat{N}_j}=N/2$, these states maximise the variance and can be considered optimal. This should come as no surprise -- under the evolution (\ref{Ugeneral}) they lead to a superposition of two different displaced motions for the mechanical elements (i.e. oscillating cat states). The enhancements with $N/2$ from each cavity (equal to the standard deviations $\Delta \hat{N}_j$) are then roughly proportional to the maximum separation of the superposition (see also Example 1. in section \ref{Sec:semiclassical}).

Note in this case, one can also readily calculate the entanglement through the von Neumann entropy of one of the reduced states, $S=-\Tr \rho_1 \ln \rho_1 = -\sum_j \lambda_j \ln \lambda_j $, where $\lambda_j$ are the eigenvalues of $\rho_1$. Under the assumptions above, it is straightforward to show, 
\begin{equation}\label{rho1}
\rho_1= \hat{U}_1 \left[ \sum_{r,l}a_r a_l^* \sum_\mu |b_\mu|^2 e^{iDv_\mu(u_r-u_l)} \ketbra{u_r}{u_l} \right] \hat{U}_1^\dagger.
\end{equation}
For a two level system, $\rho_1$ corresponds to a $2\times 2$ matrix, and so the eigenvalues are given by the well known formula,
\begin{equation}
\lambda_\pm = \frac{1}{2} \left( \tr \rho_1 \pm \sqrt{[\tr \rho_1]^2 - 4 \det \rho_1}\right),
\end{equation}
where from (\ref{rho1}), 
\begin{equation}\label{vonNeumannTrDet}
\begin{aligned}
\tr \rho_1 &= 1 ,\\
\det \rho_1 &= |a_1|^2 |a_2|^2 \left( 1 - \left[|b_1|^4 +|b_2|^4   +2 |b_1|^2 |b_2|^2 \cos D (u_1-u_2)(v_1-v_2) \right] \right) .
\end{aligned}
\end{equation}
As the entropy is maximised (but not necessarily maximal) when the separation between eigenvalues is largest, we again have the condition $\cos D (u_1-u_2)(v_1-v_2) = \pi$.

In optomechanical systems, however, generating Fock state superpositions with large $N$ is extremely difficult and a much higher photon number variance can be achieved by using coherent states. In this case, (\ref{SL}) does not extend all the way up to the point of maximum entanglement. This is already clear from the next term in (\ref{App:SLseries}), which can be expanded to arbitrary order using the identity $\avg{\alpha|\hat{N}^n|\alpha} = B_n(N_p)$, where $B_n(x)$ is a Bell polynomial. While this form can be valid for large $N_p$ (small $D$), its accuracy typically grows slowly for increasing $S_L$. An alternative expansion can be found starting from the purity (\ref{App:Purity1}) with $\ket{\psi_1}=e^{-\frac{1}{2}|\alpha|^2}\sum_{p=0}^{\infty} \frac{\alpha^p}{\sqrt{p!}}\ket{p}$ (and similarly for $\ket{\psi_2}=\ket{\beta}$). Then setting $X=D(p-q)$, with $n=0$ in the identity (\ref{App:coherentexpectationidentity}) (see below) leads to,
\begin{equation} 
\tr[\rho_1^2]= e^{-2|\alpha|^2} \sum_{p,q=0}^{\infty} \frac{|\alpha|^{2p}}{p!}\frac{|\alpha|^{2q}}{q!}e^{2|\beta|^2 [\cos{D(p-q)}-1]},
\end{equation}
which can alternatively be recast in the slightly more computationally convenient form, 
\begin{equation}\label{App:SLnumerics}
S_L= 1-e^{-2|\beta|^2} \sum_{r=0}^{\infty}\sum_{k=0}^{r} e^{2|\alpha|^2 [\cos{D(r-2k)}-1]} \frac{|\beta|^{2r}}{k!(r-k)!}.
\end{equation}
As long as $N_p$ is not too large, this allows a reasonably accurate evaluation of the linear entropy up to $S_L \sim 1$. We find that for most values, this turns out to be approximately a function of $DN_p$ only (assuming $(\Delta\hat{N}_j)^2=N_p$ is identical in each cavity). In practise, this means that we can still use (\ref{SL}) to estimate the entanglement time. A conservative choice for the measurable entanglement threshold is to assume $S_L= \mathcal{O}(1)$, where now the proportionality constant needs to be determined numerically. For concreteness we choose the point corresponding to the maximum rate of increase the linear entropy, which is found to be at approximately $|D(\tau_e)|= \frac{1}{2\sqrt{2}N_p}$, see figure \ref{fig:SL}. Beyond $\tau_e$, entanglement gains become less significant, with the linear entropy changing only very slowly above $DN_p\approx 4$.

%-------------------------------------------------
\section{Noise}\label{App:Noise}

The evolution equation (\ref{maintextmastereq}) used in section \ref{Sec:Decoherence} is the limiting case of the standard quantum Brownian motion master equation applied to two interacting systems. It can be derived from first principles by introducing a position-position coupling between the system variables and an infinite bath of harmonic oscillators \cite{Caldeira1983,HuPazZhang1992}. For a single system, this is given in operator form by \cite{Zurek2003},
\begin{equation}\label{MasterEquationGeneral}
\frac{d \rho(t)}{dt} = \frac{1}{i\hbar}[\hat{H}_{\text{ren}},\rho(t)] - \frac{i\gamma_R(t)}{\hbar}[\hat{x},\{\hat{p},\rho(t)\}] - \frac{f(t)}{\hbar}[\hat{x},[\hat{p},\rho(t)]] - h(t)[\hat{x},[\hat{x},\rho(t)]] \,,
\end{equation}  
where $\{\cdot,\cdot\}$ is the usual anticommutator and $\hat{H}_{\text{ren}}$ corresponds to the renormalised system Hamiltonian (for example, a frequency shift for a free harmonic oscillator), while the $\gamma_R(t)$, 
$f(t)$ and $h(t)$ terms describe damping and anomalous and normal diffusion, respectively. In general these coefficients are difficult to calculate explicitly, but exact forms are known in special cases. The most well studied is the high temperature limit, which is valid when $k_BT$ is the highest energy scale in the system. This nevertheless still gives good results for low temperatures, particularly in the macroscopic limit relevant here \cite{Zurek2003}. 

If the system-bath couplings are distributed according to an ohmic spectral density, with suitable cut-off, the normal diffusive term is much larger than $f(t)$ and quickly settles to a constant value $h(t) = \frac{2m k_B T \gamma_R}{\hbar^2}$. One then notes that in the macroscopic limit 
the second term in (\ref{MasterEquationGeneral}) can be ignored, leading to 
\begin{equation}\label{MasterEquationSimple}
\frac{d \rho(t)}{dt} = \frac{1}{i\hbar}[\hat{H}_{\text{ren}},\rho(t)] - \frac{2m k_B T \gamma_R}{\hbar^2}[\hat{x},[\hat{x},\rho(t)]].
\end{equation}
This equation is Markovian, and could in principle have also been derived within the Lindblad framework. For two weakly interacting systems, each with their own bath, we can expect that the indirect coupling of one system to the other's environment will be very small. In the proposal considered here, this should hold for any realistic experiment, where $\gamma_g \ll 1$, and so it is reasonable to suppose a master equation of the form (\ref{maintextmastereq}), 
\begin{equation}\label{appendixmastereq}
\frac{{d \rho(t)}}{dt} = \frac{1}{i\hbar} [\hat{H}_{\text{ren}},\rho(t)] - \frac{\Upsilon_1}{\hbar^2}[\hat x_1,[\hat x_1,\rho(t)]] -\frac{\Upsilon_2}{\hbar^2}[\hat x_2,[\hat x_2,\rho(t)]],
\end{equation}
where $\Upsilon_j=2m_jk_BT_{j}\gamma_{R,j}$ is the normal diffusion coefficient associated to each cavity, and for convenience we take $\hat{H}_{\text{ren}}=\hat{H}$. This is justified as the mechanical frequency shifts in the expected regimes will typically be very small and therefore not greatly effect the entanglement dynamics. The extension from (\ref{MasterEquationSimple}) to (\ref{appendixmastereq}) has been validated both from first principles derivations, and numerically, for simple systems, such coupled harmonic oscillations \cite{Rivas2010}, and shows good agreement in the small coupling limit. Furthermore, in the case of equal temperature baths, (\ref{appendixmastereq}) leads to essentially identical results to the strong coupling description, if positivity is enforced via the secular approximation. It should be noted that the same may not be true for the optical modes, where in particular we aim to work in the strong optomechanical coupling regime. In the present work, however, we will not address combined optical and mechanical noise.

For our purposes, it is sufficient to solve (\ref{appendixmastereq}) for the reduced cavity state alone. Here we make use of a straightforward generalisation of the approach in \cite{Adler2005}. For completeness, we repeat the relevant steps below. First, we transform to the the c.o.m. and stretch modes so that $\hat{H}$ is diagonalised. When the cavities are identical, i.e. $\Upsilon_1=\Upsilon_2$ then (\ref{appendixmastereq}) can be rewritten as,
\begin{equation}\label{mastereq}
\frac{{d \rho(t)}}{dt} = \frac{1}{i\hbar} [\hat{H},\rho(t)] - \frac{\Upsilon}{\hbar^2}[\hat x_c,[\hat x_c,\rho(t)]] - \frac{\Upsilon}{\hbar^2}[\hat x_s,[\hat x_s,\rho(t)]],
\end{equation}
where $\Upsilon=2mk_BT \gamma_R$. We then define the density matrix,
\begin{equation}\label{projectedDM}
\rho_D(t)=\bra{p,q}\rho(t)\ket{m,n},
\end{equation}
where $\ket{m,n}=\ket{m}_1\ket{n}_2$ are field states of the two optical cavities. Thus the components of the reduced cavity state are given simply by,
\begin{equation}
\rho_{p,q,m,n}^{cav}(t)=\Tr_M[\rho_D(t)].
\end{equation}
Using the projection (\ref{projectedDM}), we can re-write (\ref{mastereq}) as,
\begin{equation}
\begin{aligned}
\frac{{d \rho_D(t)}}{dt} &= -\frac{i}{\hbar} \hat H^{p,q}\rho_D(t) + \frac{i}{\hbar} \rho_D(t)\hat H^{m,n} - \frac{\Upsilon}{\hbar^2}[\hat x_c,[\hat x_c,\rho_{D}(t)]] -\frac{\Upsilon}{\hbar^2}[\hat x_s,[\hat x_s,\rho_{D}(t)]],
\end{aligned}
\end{equation}
where we have used that $\hat H$ is diagonal in the photon number basis and we defined $\hat H^{m,n}=\avg{m,n|\hat{H}|m,n}$ as the effective Hamiltonians acting on the normal mode subspace,
\begin{equation}
\begin{aligned}
\hat H^{m,n} &= \hbar \omega_0 (m+n) + \hbar \omega_c \bd_c \b_c +\hbar \omega_s \bd_s \b_s - \lambda_c^{m,n}(t)(\bd_c + \b_c) - \lambda_s^{n,m}(t)(\bd_s + \b_s),
\end{aligned}
\end{equation} 
with, 
\begin{equation}
\begin{aligned}
\lambda_c^{m,n}(t)&=\hbar \omega_c k_c(t)(m-n),\\
\lambda_s^{m,n}(t)&= \hbar \omega_s\left(-k_s(t)(m+n)+\mathcal{D}_{1,s}(t)\right).
\end{aligned}
\end{equation}

The next step is to move to the interaction picture by defining $\hat\rho_D^I(t) = \hat U^{ p,q \dagger}(t)\rho_D(t)\hat U^{m,n}(t)$, where the $\hat U^{m,n}(t)$ are defined through,
\begin{equation}
i \hbar \frac{\partial}{\partial t}\hat U^{m,n}(t)=\hat H^{m,n}\hat U^{m,n}(t)\,,
\end{equation}
and since $\hat{H}$ is photon number conserving, $\hat U^{m,n}(t) = \avg{m,n|\hat{U}(t)|m,n}$.
Thus we have,
\begin{equation}\label{mastereqI}
\begin{aligned}
\frac{{d \rho_D^I(t)}}{dt} &=  - \frac{\Upsilon}{\hbar^2}\hat U^{ p,q \dagger}(t)[\hat x_c,[\hat x_c,\rho_D(t)]]\hat U^{m,n}(t)  -\frac{\Upsilon}{\hbar^2}\hat U^{ p,q \dagger}(t)[\hat x_s,[\hat x_s,\rho_D(t)]]\hat U^{m,n}(t),
\end{aligned}
\end{equation}
We now multiply from the left and right by $\hat U^{p,q}(u)$ and $\hat U^{m,n \dagger}(u)$, respectively and take the partial trace over the mirror modes, 
	\begin{equation}\label{mastereqITr}
	\begin{aligned}
	\Tr_M\left\{\hat U^{p,q}(u) \frac{{d \rho_D^I(t)}}{dt} \hat U^{m,n \dagger}(u)\right\} &=  -\frac{\Upsilon}{\hbar^2}\Tr_M\left\{\hat U^{ p,q \dagger}(t-u)[\hat x_c,[\hat x_c,\rho_D(t)]]\hat U^{m,n}(t-u) \right\}\\
	& \quad -\frac{\Upsilon}{\hbar^2}\Tr_M\left\{\hat U^{ p,q \dagger}(t-u)[\hat x_s,[\hat x_s,\rho_D(t)]]\hat U^{m,n}(t-u)\right\} \\
	&=   - \frac{\Upsilon}{\hbar^2}\Tr_M\left\{[\hat x_c,[\hat x_c,\hat U^{m,n}(t-u)\hat U^{ p,q \dagger}(t-u)]]\rho_D(t) \right\}\\
	&  \quad -\frac{\Upsilon}{\hbar^2}\Tr_M\left\{[\hat x_s,[\hat x_s,\hat U^{m,n}(t-u)\hat U^{ p,q \dagger}(t-u)]]\rho_D(t)\right\},
	\end{aligned}
	\end{equation}
where we have used the cyclic property of the trace in the second line. The double commutators can be evaluated using
\begin{equation}
[\hat x_\mu,[\hat x_\mu,\hat U^{m,n}(t)\hat U^{ p,q \dagger}({t})]]= [\hat {\tilde{x}}_\mu^2(t)+\hat {\tilde{x}}_\mu^2(0) - 2\hat{\tilde{x}}_\mu(0)\hat{\tilde{x}}_\mu(t)]\hat U^{m,n}(t)\hat U^{ p,q \dagger}(t),
\end{equation}
where,
\begin{equation}
\hat{\tilde{x}}_\mu(t)=\hat{\tilde{U}}^\dagger(t)\hat x_\mu \hat{\tilde{U}}(t),
\end{equation}
and,
\begin{equation}
\begin{aligned}
\hat{\tilde{U}}(t)&=\hat U^{p,q}(t)\hat U^{ m,n \dagger}(t) \\
&=e^{-i(\omega_0t - A)(p+q-m-n)}e^{iB(p^2+q^2-m^2-n^2)}e^{iD(pq-mn)} \times \\
& \quad \times \hat D_c[-  e^{-i\omega_c t}  K_{\hat{N}_c}(q-p+m-n)]\hat D_s[ -   e^{-i\omega_s t}  K_{\hat{N}_s}(p+q-m-n)],
\end{aligned}
\end{equation}
The first set of phase factors do not contribute to $\hat{\tilde{x}}_\mu(t)$, and so we find, 
\begin{equation}
\begin{aligned}
\hat{\tilde{x}}_c(t) &= \hat{\tilde{x}}_c(0)  - 2x_{0,c}\Re[e^{-i\omega_c t} K_{\hat{N}_c}](q-p+m-n), \\
\hat{\tilde{x}}_s(t) &= \hat{\tilde{x}}_s(0)  - 2x_{0,s}\Re[e^{-i\omega_s t} K_{\hat{N}_s}](q+p-m-n),
\end{aligned}
\end{equation}
and so the double commutators become,
\begin{equation}
\begin{aligned}
[\hat x_c,[\hat x_c,\hat U^{m,n}(t)\hat U^{ p,q \dagger}({t})]]&= 4x_{0,c}^2\Re[e^{-i\omega_c t} K_{\hat{N}_c}]^2(q-p+m-n)^2\hat U^{m,n}(t)\hat U^{ p,q \dagger}(t), \\
[\hat x_s,[\hat x_s,\hat U^{m,n}(t)\hat U^{ p,q \dagger}({t})]]&= 4x_{0,s}^2\Re[e^{-i\omega_s t} K_{\hat{N}_s}]^2(q+p-m-n)^2 \hat U^{m,n}(t)\hat U^{ p,q \dagger}(t)\,.
\end{aligned}
\end{equation}
Substituting into (\ref{mastereqITr}) we find,
	\begin{equation}\label{mastereqITr2}
	\begin{aligned}
	\frac{d}{dt}\Tr_M\left\{\hat U^{p,q }(u)  \rho_D^I(t) \hat U^{m,n\dagger}(u)\right\} &=   -\Gamma(t,u)\Tr_M\left\{\hat U^{ p,q }(u) \rho_D^I(t) \hat U^{m,n\dagger}(u)\right\},
	\end{aligned}
	\end{equation}
where,
\begin{equation}
\begin{aligned}
\Gamma(t,u)&= \frac{4\Upsilon}{\hbar^2}x_{0,c}^2 \Re[e^{-i\omega_c (t-u)} K_{\hat{N}_c}(t-u)]^2(q-p+m-n)^2\\ 
&   \quad + \frac{4\Upsilon}{\hbar^2}x_{0,s}^2 \Re[e^{-i\omega_\mu (t-u)} K_{\hat{N}_s}(t-u)]^2(q+p-m-n)^2 ,
\end{aligned}
\end{equation}
This can be solved by a simple integration, and so (setting $u=t$) we find, 
\begin{equation}\label{App:rhocavsol}
\begin{aligned}
\rho_{p,q,m,n}^{cav}(t) &= e^{\left.-\int_0^t \Gamma(s,u)ds\right|_{u=t}} \Tr_M\left\{\hat U^{ p,q }(t)\rho_D(0) \hat U^{m,n\dagger}(t)\right\}\\
&= e^{\left.-\int_0^t \Gamma(s,u)ds\right|_{u=t}}  \Tr\left\{\hat U^{ p,q }(t)\rho_{M}(0) \hat U^{m,n\dagger}(t)\right\}\otimes \bra{p,q}\rho_{cav}(0)\ket{m,n},
\end{aligned}
\end{equation}
where $\rho_D^I(0)=\rho_D(0)=\bra{p,q}\rho(0)\ket{m,n}$.
This leads to an effective heating of the mechanical modes and a thermal contribution to reduced cavity state which no longer vanishes at the respective decoupling times of the optical and mechanical states. To see this explicitly we note from Appendix \ref{App:RhoCavity} that under unitary evolution, the components of the (reduced) cavity state when the mechanics is initially in a thermal state are given by, 
\begin{equation}
\begin{aligned}
\rho_{p,q,m,n}^{cav}(t)&=a_p a_m^* b_q b_n^* e^{-i(\omega- A)(p+q-m-n)t} e^{iB(m^2+n^2-p^2-q^2)}e^{iD(pq-mn)} e^{-\frac{1}{2}|K_{\hat{N}_c}|^2(\bar{n}_s+\frac{1}{2})(q-p+m-n)^2 }\\
& \quad \times e^{-\frac{1}{2}\left(|K_{\hat{N}_s}|^2(\bar{n}_s+\frac{1}{2})(p+q-m-n)^2  - (C_sK_{\hat{N}_s}^*-C_s^*K_{\hat{N}_s})(p+q-m-n)\right) } \,.
\end{aligned}
\end{equation}
for arbitrary initial cavity states $\ket{\psi_1(0)}=\sum a_m \ket{m}$, $\ket{\psi_2(0)}=\sum b_m \ket{m}$. 
This implies we can formally replace the thermal occupation $\bar{n}_\mu$ by a time dependent term, 
\begin{equation}\label{eq:barnt}
	\bar{n}_\mu(t)=\bar{n}_\mu(0)+\frac{\kappa_\mu^\mathrm{dec}(t)}{|K_{\hat{N}_\mu}|^2}\,,
\end{equation}
where,
\begin{equation}\label{eq:kappatdef}
\kappa_\mu^\mathrm{dec}(t)= \int_0^t \frac{8\Upsilon}{\hbar^2}x_{0,\mu}^2\Re[e^{-i\omega_\mu (s-t)} K_{\hat{N}_\mu}(s-t)]^2 ds .
\end{equation}
For the modulation frequency $\Omega_n=(1-1/n)(\omega_c+\omega_s)$ and $\phi_1=\pi/2$, $\kappa_\mu^\mathrm{dec}(t_q)$ can be evaluated to zeroth order in $\gamma_g$ as,
\begin{equation}\label{kappadec}
\begin{aligned}
\kappa_\mu^\mathrm{dec}(t_q) &= \frac{ 4\Upsilon}{\hbar^2}x_{0,m}^2 \frac{qn\pi}{\omega_m} \left(3 k_0^2 + \epsilon^2 \frac{n^2(1+(2n-1)^2) }{2(2n-1)^2}\right)\\
& \approx  \frac{ \Upsilon x_{0,m}^2}{\hbar^2\omega_m \gamma_g } D(t_q),
\end{aligned}
\end{equation}
for $\mu=c,s$.

\section{Evaluation of Determinant-based Entanglement Witnesses for Continuous Variable States}\label{App:Witness}

In section \ref{Sec:Witness} it was noted that the idealised limiting state of the final (entangled) optical fields will be of the form $(1+i)\ket{\alpha_1,\alpha_2}+(1-i)\ket{-\alpha_1,-\alpha_2}$. While many of the simplest continuous variable criteria fail to identify entanglement in states of this form, Shchukin and Vogel \cite{ShchukinVogel2005} introduced an infinite series of inequalities for which the violation of any is sufficient to witness entanglement. These can be conveniently written as the determinant of a matrix of moments of the state, where negativity implies the violation of one set of inequalities. The simplest which is suitable for our purposes is given by, 
\begin{equation}\label{Wit1}
\mathcal{W}_1(t) = \begin{vmatrix}
1 & \braket{a_2^{\dagger}} & \braket{a_1a_2^{\dagger}}  \\ 
\braket{a_2} & \braket{a_2^{\dagger}a_2} & \braket{a_1a_2^{\dagger}a_2}  \\ 
\braket{a_1^{\dagger}a_2} & \braket{a_1^{\dagger}a_2^{\dagger}a_2} &\braket{a_1^{\dagger}a_1a_2^{\dagger}a_2}  
\end{vmatrix}.
\end{equation}
For convenience we move to the Heisenberg picture, $\hat{O}(t)=\hat{U}^\dagger\hat{O}\hat{U}$, where in the symmetric cavity limit the evolution operator is given explicitly by (\ref{Ugeneral}) and (\ref{eq:CoefficientsRed}),
\begin{equation}
\begin{aligned}
U(t)&= e^{i(-\omega_0t+A})(\hat N_1 + \hat N_2) e^{-i\omega_c \hat b_c^{\dagger}\hat b_ct}e^{-i\omega_s \hat b_s^{\dagger}\hat b_st}   e^{i\theta} e^{iB(\hat N_1^2 +\hat N_2^2)} e^{iD \hat N_1\hat N_2}  \times\\
& \quad \times D_c[-K_{\hat{N}_c}(\hat N_2-\hat N_1)]D_s[C_s - K_{\hat{N}_s}(\hat N_1+\hat N_2)] ,
\end{aligned}
\end{equation}
with $\omega_{s}=\omega_m\sqrt{1 - 4\gamma_g}$ and $\omega_{c}=\omega_m$. Using the relation, 
\begin{equation}
D(C  - K(\hat{N}_2+(-1)^k \hat{N}_1))=D(C)D( - K\hat{N}_2)D((-1)^{k+1} K\hat{N}_1) e^{ + \frac{1}{2}(CK^*-C^*K)(\hat{N}_2+(-1)^k\hat{N}_1)},
\end{equation}
Together with $D_\mu^\dagger(X\hat{N}_j)\hat{a}_jD_\mu(X\hat{N}_j) = D_\mu(X)\hat{a}_j$, and $e^{-X\hat{N}^2}\hat{a}e^{X\hat{N}^2}=e^{X(2\hat{N}+1)}\hat{a}$ we find,
\begin{equation}
\begin{aligned}\label{timedependentmoments}
\hat{a}_1(t)&= e^{i\phi_0}\hat{\mathcal{B}}_1 e^{i(D\hat{N}_2 + 2B\hat{N}_1)} \hat{a}_1\,,\\
\hat{a}_2(t)&= e^{i\phi_0} \hat{\mathcal{B}}_2e^{i(D\hat{N}_1 + 2B\hat{N}_2)} \hat{a}_2\,,\\
(\hat{a}_1\hat{a}^\dagger_2)(t)&=
\hat{\mathcal{B}}_{12}\hat{a}^\dagger_2e^{i(2B-D)(\hat{N}_1-\hat{N}_2)} \hat{a}_1\,, 
\\
(\hat{a}^\dagger_j\hat{a}_j)(t)& = \hat{a}^\dagger_j\hat{a}_j\,,\\
(\hat{a}_1\hat{a}^\dagger_2\hat{a}_2)(t)& =\hat{a}_1(t) \hat{a}^\dagger_2\hat{a}_2 = e^{iD}\hat{a}^\dagger_2\hat{a}_1(t) \hat{a}_2\,.
\end{aligned}
\end{equation}
Where $\phi_0=-\omega_0t + A +B$ and,
\begin{equation}
\begin{aligned}
\hat{\mathcal{B}}_j&=e^{+\frac{1}{2}(C_sK_{\hat{N}_s}^*-C_s^*K_{\hat{N}_s})}D_c(-(-1)^j K_{\hat{N}_c})D_s(-K_{\hat{N}_s})\,, \\
\hat{\mathcal{B}}_j^2&=e^{+(C_sK_{\hat{N}_s}^*-C_s^*K_{\hat{N}_s})}D_c(-(-1)^j 2K_{\hat{N}_c})D_s(-2K_{\hat{N}_s})\,, \\
\hat{\mathcal{B}}_{12}&=\hat{\mathcal{B}}_1\hat{\mathcal{B}}_2^\dagger=D_c(+2K_{\hat{N}_c})\,,\\
\hat{\mathcal{B}}_1\hat{\mathcal{B}}_2 & =e^{+(C_sK_{\hat{N}_s}^*-C_s^*K_{\hat{N}_s})}D_s(-2K_{\hat{N}_s}) \\
\end{aligned}
\end{equation}
are operators acting on the mechanical modes. Here we consider two scenarios: 1) The mechanical modes initially in a coherent state $\beta_\mu$, and 2) The mechanical modes initially in a thermal state  $\rho_{\rm{th},\mu}$. For each case the expectation value of the displacement operator is given by,
\begin{equation}
\begin{aligned}\label{AvgD}
\avg{D_\mu(X)}_\beta & = e^{-\frac{1}{2}|X|^2}e^{X\beta^* - X^*\beta}\,,\\
\avg{D_\mu(X)}_{\rho_{th}} & =e^{-\frac{1}{2}|X|^2} e^{-(\Re[X]^2+\Im[X]^2)\bar{n}_\mu}=e^{-|X|^2(\bar{n}_\mu+\frac{1}{2})}\,.
\end{aligned}
\end{equation} 
So we have,
\begin{equation}\label{BetaAvgscoherent}
\begin{aligned}
\avg{\mathcal{B}_j}_\beta&=e^{-\frac{1}{2}\left(|K_{\hat{N}_c}|^2+|K_{\hat{N}_s}|^2\right)} e^{+\frac{1}{2}(C_sK_{\hat{N}_s}^*-C_s^*K_{\hat{N}_s})} e^{(-1)^j \tilde{\beta}_c +\tilde{\beta}_s} ,\\
\avg{\mathcal{B}_j^2}_\beta&=e^{-2\left(|K_{\hat{N}_c}|^2+|K_{\hat{N}_s}|^2\right)} e^{+(C_sK_{\hat{N}_s}^*-C_s^*K_{\hat{N}_s})} e^{2[(-1)^j \tilde{\beta}_c +\tilde{\beta}_s]}, \\
\avg{\mathcal{B}_{12}}_\beta & =  e^{-2|K_{\hat{N}_c}|^2}e^{-2\tilde{\beta}_c},\\
\avg{\mathcal{B}_1\mathcal{B}_2}_\beta &= e^{-2|K_{\hat{N}_s}|^2} e^{+(C_sK_{\hat{N}_s}^*-C_s^*K_{\hat{N}_s})}e^{2\tilde{\beta}_s} ,
\end{aligned}
\end{equation}
 where $\tilde{\beta}_\mu= -(K_{\hat{N}_\mu}\beta_\mu^* - K_{\hat{N}_\mu}^*\beta_\mu)$ and,
\begin{equation}\label{BetaAvgsthermal}
\begin{aligned}
\avg{\mathcal{B}_j}_{\rho_{th}}&=e^{-\frac{1}{2}\left(|K_{\hat{N}_c}|^2(\bar{n}_c+\frac{1}{2})+|K_{\hat{N}_s}|^2(\bar{n}_s+\frac{1}{2})\right)} e^{+\frac{1}{2}(C_sK_{\hat{N}_s}^*-C_s^*K_{\hat{N}_s})}, \\
\avg{\mathcal{B}_j^2}_{\rho_{th}}&=e^{-2\left(|K_{\hat{N}_c}|^2(\bar{n}_c+\frac{1}{2})+|K_{\hat{N}_s}|^2(\bar{n}_s+\frac{1}{2})\right)} e^{+(C_sK_{\hat{N}_s}^*-C_s^*K_{\hat{N}_s})},\\
\avg{\mathcal{B}_{12}}_{\rho_{th}} & =  e^{-2|K_{\hat{N}_c}|^2(\bar{n}_c+\frac{1}{2})},\\
\avg{\mathcal{B}_1\mathcal{B}_2}_{\rho_{th}} &= e^{-2|K_{\hat{N}_s}|^2(\bar{n}_s+\frac{1}{2})} e^{+(C_sK_{\hat{N}_s}^*-C_s^*K_{\hat{N}_s})},
\end{aligned}
\end{equation}

The final ingredient are the expectation values with respect to the cavity states, which we take to be coherent with parameters $\alpha_1$ and $\alpha_2$ for the two cavities, respectively. 
Using
\begin{equation}\label{App:coherentexpectationidentity}
\avg{e^{iX\hat{N}_j} \hat{a}_j^n}_{\alpha_j}=\alpha_j^n e^{-|\alpha_j|^2 \left(1-e^{iX}\right)},
\end{equation}   
we have,
\begin{equation}
\begin{aligned}\label{timedependentmomentsav}
\avg{\hat{a}_1(t)} &=\alpha_1 e^{i\phi_0} \avg{\mathcal{B}_1} e^{-|\alpha_1|^2\left(1-e^{2iB}\right)}e^{-|\alpha_2|^2\left(1-e^{iD}\right)},\\
\avg{\hat{a}_2(t)}  &=\alpha_2 e^{i\phi_0} \avg{\mathcal{B}_2} e^{-|\alpha_2|^2\left(1-e^{2iB}\right)}e^{-|\alpha_1|^2\left(1-e^{iD}\right)},\\
\avg{(\hat{a}_1\hat{a}^\dagger_2)(t)} & = \alpha_1 \alpha_2^*  \avg{\mathcal{B}_{12}} e^{-|\alpha_1|^2\left(1-e^{-i(D-2B)}\right)}e^{-|\alpha_2|^2\left(1-e^{i(D-2B)}\right)},\\
\avg{(\hat{a}_j^\dagger\hat{a}_j)(t)} & = |\alpha_j|^2,\\
\avg{(\hat{a}_1\hat{a}^\dagger_2\hat{a}_2)(t)}&=\alpha_1 |\alpha_2|^2e^{iD} e^{i\phi_0} \avg{\mathcal{B}_1} e^{-|\alpha_1|^2\left(1-e^{2iB}\right)}e^{-|\alpha_2|^2\left(1-e^{iD}\right)},\\
\avg{(\hat{a}_1^\dagger\hat{a}_1\hat{a}^\dagger_2\hat{a}_2)(t)}&=|\alpha_1 |^2|\alpha_2|^2.
\end{aligned}
\end{equation}

These expressions are generally too complicated to gain much insight. However, by considering the form of the determinant (\ref{Wit1}) we observe that the phase $\phi_0$ ultimately vanishes together with the imaginary exponential terms in (\ref{BetaAvgscoherent}) and (\ref{BetaAvgsthermal}). Similarly, only the modulus of $\alpha_1$ and $\alpha_2$ is important  and so from symmetry of the set-up this implies that the optimal choice of initial cavity states is $\alpha_1=\alpha_2 \equiv \alpha$ and $|\alpha|^2=N_p$. With these restrictions, the determinant witness (\ref{Wit1}) can be written as, 
\begin{equation}\label{appWit1full}
\begin{aligned}
\mathcal{W}_1(t_q) &= N_p^3 \left( 1 - e^{-4\kappa_c^\mathrm{th}  -4N_p(1-\cos{(D-2B)})}  - 2 e^{-\kappa_c^\mathrm{th} -\kappa_s^\mathrm{th}  - 2N_p (2-\cos D-\cos2B)} \right.\\
& \quad \left.+ 2 e^{-3\kappa_c^\mathrm{th}  -\kappa_s^\mathrm{th}  - 2N_p(3-\cos D-\cos{2B}-\cos{(D-2B)})}\cos D \right) ,
\end{aligned}
\end{equation}
where $\kappa_{\mu}^\mathrm{th} = |K_{\hat{N}_{\mu}}|^2(\bar{n}_{\mu}+1/2)$.
	
We note, however, that the $B$ terms should not in principle contribute to the entanglement, even though they have a non-trivial effect on the witness. An analogous situation arises in the estimation of the gravitational field by a single optomechanical cavity, where it is known that the quantum Fisher information (QFI) is independent of the corresponding $B$ coefficient at the decoupling time, however certain POVMs only saturate the QFI when this is an integer multiple of $2 \pi$ \cite{Qvarfort2021}. This is occurs when using homodyne measurements, for example, and typically require $k_0 \in \mathbb{Z}$. Interestingly, we find from a numerical analysis, that the optimal values of $B(t_q)$ for the witness are integer multiples of $\pi$. Considering a measurement time $t_q=qn \pi/(1/\omega_c + 1/\omega_s)$, with the optomechanical coupling modulated at frequency $\Omega_n=\frac{1}{2}(1-1/n)(\omega_c+\omega_s)$, we find from $B=B_c+B_s$ that,
\begin{equation}
B(t_q)=2qn\pi(1+3\gamma_g)\left(k_0^2 +\frac{n^2}{2(2n-1)}\epsilon^2\right).
\end{equation}
Thus in the large $n$ limit, we also see that approximately integer values of $k_0$ and $\epsilon$ are favourable\footnote{On the other hand, heterodyne measurements achieve the same scaling as the QFI (up to a factor of $2$)  independent of this constraint, and so we may hypothesize that a heterodyne-based witness could be more convenient from an experimental perspective.}. Adopting these $B(t_q)$ values leads to immediately to the witness in the main text (\ref{maintextWitSimple}).
The prefactor (generically $N_p^3$) should not be attributed to the entanglement scaling estimated in section \ref{ModulatedEntanglementTime}. Instead this term acts like a signal gain, which should be offset against a proper noise analysis of measurements of the moments used in (\ref{Wit1}). 

When the initial mode temperatures are equal, we find that $\kappa_c^\mathrm{th}(t_q)=\kappa_s^\mathrm{th}(t_q)$ to first order in $\gamma_g$,
\begin{equation}\label{App:Kapprox}
\begin{aligned}
\kappa_{\mu}^\mathrm{th}(t_q) & \approx \frac{ (2 \pi n \gamma_g)^2 \left( k_0^2 (2n-1)^2 + \left(n-1\right)^2 n^2 \epsilon^2\right)}{2(2n-1)^2} (\bar{n}_{\mu}+1/2)\\
& \approx \frac{1}{2}(\pi n \gamma_g)^2  \left( 4 k_0^2 +  n^2 \epsilon^2\right)(\bar{n}_{\mu}+1/2),
\end{aligned}
\end{equation}
where in the second line we have assumed $n$ large. Thus,
\begin{equation}\label{app:witsimp}
	\mathcal{W}_1(t_q)= 4N_p^3 e^{-4y}\left(e^{2y}\sinh^2(y)-z^2\right),
\end{equation}
where $y=2N_pz^2+\kappa^\mathrm{th}(t_q)$ and $z=\sin(D(t_q)/2)$. Taking the derivative with respect to $z$, the minimum will satisfy (\ref{WitnessTPcondition0}), 
\begin{equation}\label{App:WitnessTPcondition0}
4z_{\min}^2+e^{2y_{\min}}=1+\frac{2z_{\min}}{4N_pz_{\min}+\kappa^{\mathrm{th} \prime}|_{z_{\min}}}.
\end{equation}
For $\kappa^{\mathrm{th}\prime}|_{z_{\min}} \approx 0$, a formal solution for $z_{\min}$ can be written as,
\begin{equation}
z_{\min}=\frac{ \sqrt{1+2N_p-2W(e^{\frac{1}{2}+N_p+2\kappa^\mathrm{th}}N_p)}}{2\sqrt{2N_p}},
\end{equation}
where $W(x)$ is the Lambert-$W$ function defined as the solution to $W(x)e^{W(x)}=x$. For large $x$, we can approximate $W(x)$ asymptotically as,
\begin{equation}
W(x)=\ln\left(\frac{x}{W(x)}\right) \approx \ln\left(\frac{x}{\ln {x}}\right),
\end{equation}
then to first order in $1/N_p$, the term $W(e^{\frac{1}{2}+N_p+2\kappa^\mathrm{th}}N_p) = \frac{1}{2} + N_p -\frac{1}{2N_p}[1-4(N_p-1)\kappa^{\rm{th}}]$, and so
\begin{equation}
z_{\min} \approx\frac{\sqrt{1-4(N_p-1)\kappa^\mathrm{th}}}{2\sqrt{2}N_p} \approx\frac{\sqrt{1-4N_p\kappa^\mathrm{th}}}{2\sqrt{2}N_p},
\end{equation}
which agrees with the simple analysis provided in the main text.

When $\kappa'|_{z_{\min}}$ is significant, it is convenient to first write $\kappa$ explicitly as a function of $z$. For $\epsilon=0$, $\kappa^\mathrm{th}(z) \approx \Gamma^{\rm{th}}_0 z^2$, with $\Gamma_0^{\rm{th}}=\frac{2\bar{n}+1}{36k_0^2}$ (see (\ref{kappathermalz})), which immediately leads to (\ref{Dconstopt}) under the substitutions $N_p\rightarrow N_p+\Gamma_0^{\rm{th}}/2$ and $\kappa^\mathrm{th}\rightarrow0$. However, for modulated coupling, where $\kappa^{\rm{th}}=\Gamma z^{4/3}$, finding a formal solution to the minimum condition is difficult, and so we instead resort to an expansion of (\ref{App:WitnessTPcondition0}) under the assumption $z_{\min},N_p z_{\min}^2\ll 1$, and $N_p \gg1$ (which are all consistent with the expected experimental regimes). In order to preserve the correct behaviour as $\kappa^{\rm{th}}\rightarrow 0$, it is convenient to first multiply (\ref{App:WitnessTPcondition0}) by $(3/2)(4N_p z_{\min}+\kappa'|_{z_{\min}})/z_{\min}$, before expanding the resulting expression to second order in $z_{\min}$. A slight rearrangement leads to,
\begin{equation}\label{modminexpansion}
 4(6 N_p^2 +\Gamma^3)w^3  + 20N_p\Gamma w^2 +4 \Gamma^2 w -3=0,
\end{equation}
where $\Gamma = \frac{1}{2}(\bar{n} +\frac{1}{2})\left(\frac{\pi q \gamma_g}{\epsilon}\right)^{\frac{2}{3}}$, and we have made the substitution $z_{\min}\rightarrow w^{3/2}$. The solution for $w$ up to first order in $1/N_p$ is given by,
\begin{equation}
w=\frac{1-v}{2N_p^{2/3}}, \qquad \text{where } v=\frac{5\Gamma}{9N_p^{1/3}} = \frac{5}{18}\frac{\bar{n} +\frac{1}{2}}{N_p^{1/3}}\left(\frac{\pi q \gamma_g}{\epsilon}\right)^{\frac{2}{3}}.
\end{equation}
Transforming back to $z_{\min}$ leads to (\ref{Modulatedz}) in the main text. This corresponds to an optimal entanglement time of, 
\begin{equation}
t_1= \left( \frac{ 2\sqrt{2}\pi^2}{N_p \omega_m^3  \epsilon^2 \gamma_g}  \right)^{\frac{1}{3}} \sqrt{1-v}.
\end{equation} 
We now expand the exponential terms in $\mathcal{W}_1(t_1)$ in powers of $1/N_p$, ignoring overall terms proportional to any inverse powers of $N_p$ (which is consistent with our final bound on $\Gamma$). This leads to,
\begin{equation}\label{WitMinGammaMod}
\mathcal{W}_1(z_{\min}) \approx \frac{3}{8}-\frac{N_p}{4} + \frac{1}{2}N_p^{2/3}\Gamma - \frac{4}{9}N_p^{1/3}\Gamma^2 +\frac{35 \Gamma^3}{162}.
\end{equation}
In the limit $N_p \gg 1$ the first term can be neglected, and so we again recover the asymptotic noiseless witness minimum of $-N_p/4$. The solution for $\Gamma$ such that (\ref{WitMinGammaMod}) is half the optimal value is then approximately,
\begin{equation}
\Gamma \lesssim 0.33 N_p^{1/3}.
\end{equation}

%----------------------------------------------------------------
\subsection{Decoherence}\label{App:witness_decoherence}
The comparatively slower scaling of the temperature bound is a direct result of measuring close to (or at) the cavity-mechanics decoupling time. This highlights the advantage of the fractional frequencies introduced in \cite{Qvarfort2021}. On the other hand mechanical noise depends on integrating over the full motion and therefore always contains a term independent of $\gamma_g$. In Appendix \ref{App:Noise} we saw that thermal decoherence effects can be formally included by rewriting expectation values of the mechanical mode operators as (see equation \eqref{eq:barnt}),
\begin{equation}\label{BetaAvgsNoise}
\begin{aligned}
\avg{\mathcal{B}_j}_{\rho_{th}}&=e^{-\frac{1}{2}\left(|K_{\hat{N}_c}|^2(\bar{n}_c+\frac{1}{2})+|K_{\hat{N}_s}|^2(\bar{n}_s+\frac{1}{2}) +\kappa_c^\mathrm{dec}+\kappa_s^\mathrm{dec}\right)} e^{+\frac{1}{2}(C_sK_{\hat{N}_s}^*-C_s^*K_{\hat{N}_s})}, \\
\avg{\mathcal{B}_j^2}_{\rho_{th}}&=e^{-2\left(|K_{\hat{N}_c}|^2(\bar{n}_c+\frac{1}{2})+|K_{\hat{N}_s}|^2(\bar{n}_s+\frac{1}{2}) +\kappa_c^\mathrm{dec}+\kappa_s^\mathrm{dec}\right)} e^{+(C_sK_{\hat{N}_s}^*-C_s^*K_{\hat{N}_s})},\\
\avg{\mathcal{B}_{12}}_{\rho_{th}} & =  e^{-2|K_{\hat{N}_c}|^2(\bar{n}_c+\frac{1}{2})}e^{-2\kappa_c^\mathrm{dec}},\\
\avg{\mathcal{B}_1\mathcal{B}_2}_{\rho_{th}} &= e^{-2|K_{\hat{N}_s}|^2(\bar{n}_s+\frac{1}{2})} e^{-2\kappa_s^\mathrm{dec}}e^{+(C_sK_{\hat{N}_s}^*-C_s^*K_{\hat{N}_s})},
\end{aligned}
\end{equation}
where $\kappa_\mu^\mathrm{dec}(t)$ is defined in equation (\ref{eq:kappatdef}).
We find, 
\begin{equation}\label{appWit1noise}
\mathcal{W}_1(t_1) = N_p^3 \left( 1 - e^{-4\kappa_c -4N_p(1-\cos D)}  - 2 e^{-\kappa_c-\kappa_s - 2N_p (1-\cos D)} + 2 e^{-3\kappa_c -\kappa_s - 4N_p(1-\cos D)}\cos D \right) \,,
\end{equation}
where $\kappa_\mu=\kappa_\mu^\mathrm{dec}+\kappa_\mu^\mathrm{th}$. To zeroth order in $\gamma_g$, $\kappa_c^\mathrm{dec}(t_q) \approx \kappa_s^\mathrm{dec}(t_q)\gg \kappa_s^\mathrm{th}(t_q)$ and so from (\ref{kappadec}) we can set $\kappa_{\mu}(t_q)=\kappa(z(t_q))\approx\Gamma_{\rm{dec}}z(t_q)$, with $\Gamma_{\rm{dec}}=\frac{ 2\Upsilon x_{0,m}^2}{\hbar^2\omega_m \gamma_g } $. In this case the analysis proceeds in a similar fashion to the above. Substituting for $\kappa(z)$, we find the analogue of (\ref{modminexpansion}) as, 
\begin{equation}
2(12 N_p^2 + \Gamma_{\rm{dec}}^4 +12 N_p(1+\Gamma_{\rm{dec}}^2))z_{\min}^2 + 3\Gamma_{\rm{dec}}(6N_p +\Gamma_{\rm{dec}})z_{\min}+ 3(\Gamma_{\rm{dec}}-1)=0.
\end{equation}
Expanding to first orders in $1/N_p$, followed by $\Gamma_{\rm{dec}}$, the solution is given by,
\begin{equation}
z_{\min}=\frac{1}{2\sqrt{2}N_p}\left(1-\frac{3}{2\sqrt{2}}\Gamma_{\rm{dec}}\right),
\end{equation}
while to second order in $\Gamma_{\rm{dec}}$, the corresponding value of the witness is,
\begin{equation}
\mathcal{W}_1(z_{\min})=-\frac{N_p}{4}\left(1-2 \sqrt{2}\Gamma_{\rm{dec}} + \frac{5}{2}\Gamma_{\rm{dec}}^2\right).
\end{equation}
Now, solving $\min\mathcal{W}_{1} = \min a\mathcal{W}_{1|_{\Gamma_{\rm{dec}}=0}} $ gives,
\begin{equation}
\Gamma_{\rm{dec}} \leq \frac{\sqrt{2}}{5}\left(2-\sqrt{5a -1}\right),
\end{equation}
so for $a=1/2$, $\Gamma_{\rm{dec}} \lesssim 0.219$, which is typically accurate to within a little over $1\%$ of the (numerical) exact bound. In contrast, the absolute bound for this decoherence model (i.e. when infinitesimal negative values of $\mathcal{W}_1$ can be measured) is given by $\Gamma_{\rm{dec}}\leq 1$. We then have,
\begin{equation}
\gamma_R k_BT\lesssim \begin{cases}
\frac{\hbar G m}{2d ^3} ,\qquad\qquad & \text{Verify \textit{any} entanglement} \\
0.11 \frac{\hbar G m}{d ^3}, & \text{Achieve half the minimum possible value of the witness.}
\end{cases} 
\end{equation}

%---------------------------------
\section{Logarithmic negativity for optical Fock state superpositions }\label{App:RhoCavity}
In the case of finite dimensional bipartite systems, a convenient entanglement measure for mixed states is given by the logarithmic logarithmic negativity,
\begin{equation}
LN(\rho)=\log ||\rho^{T_2}||_1,
\end{equation}
where $\rho^{T_2}$ is the partial transposition of $\rho$ with respect to subsystem one and $||\rho||_1=\Tr \sqrt{\rho^\dagger \rho}$ denotes the trace norm \cite{Plenio2005}. Here we compare the entanglement behaviour when the cavity fields are prepared in equally weighted superpositions of Fock states, $\ket{\psi_{1,2}}=\frac{1}{\sqrt{2}}(\ket{0}+\ket{N})$. It is useful to first give the general expression for the time dependent reduced joint cavity state $\rho_{cav}=\Tr_{s,c}[\rho]$ when the mechanical modes are initially in the thermal states $\rho_{s,th} \otimes \rho_{c,th}$. Writing the light fields (which we assume to be pure) in the number basis, $\ket{\psi}_1=\sum a_m \ket{m}$ and $\ket{\psi}_2=\sum b_n \ket{n}$, the reduced density matrix after the evolution (\ref{Ugeneral}) can be found using (\ref{AvgD}) and the composition of displacement operators $\hat{D}(\alpha)\hat{D}(\beta)= e^{\frac{1}{2}(\alpha \beta^*-\alpha^*\beta)} \hat{D}(\alpha+\beta)$. This is given by,
\begin{equation}\label{thermaldensity}
\begin{aligned}
\rho_{cav}\left(t\right)&=\sum_{m,n,p,q} M_{mpnq} \ketbra{m}{p}\ketbra{n}{q},
\end{aligned}
\end{equation}
where,
\begin{equation}\label{thermaldensitycoeff}
\begin{aligned}
M_{mpnq}&= \,a_m a_p^* b_n b_q^* e^{-i(\omega_0 t-A)(m+n-p-q)} e^{iB(m^2+n^2-p^2-q^2)}e^{iD(mn-pq)} e^{-\frac{1}{2}|K_{\hat{N}_c}|^2(\bar{n}_{c}+\frac{1}{2})(n-m+p-q)^2 }\\
& \quad \times e^{-\frac{1}{2}\left(|K_{\hat{N}_s}|^2(\bar{n}_s+\frac{1}{2})(m+n-p-q)^2  - (C_sK_{\hat{N}_s}^*-C_s^*K_{\hat{N}_s})(m+n-p-q)\right) }, 
\end{aligned}
\end{equation}
and $\bar{n}_\mu= 1/(e^{\hbar\omega_\mu/k_B T}-1)$ is the average thermal occupation of the transformed mechanical modes. Note that as entanglement is invariant under local operations, we can equivalently consider the state $\tilde{\rho}_{cav}(t)$ defined by the coefficients, 
\begin{equation}\label{thermaldensityredcoeff}
\begin{aligned}
\tilde{M}_{mpnq}&= \,a_m a_p^* b_n b_q^* e^{iD(mn-pq)} e^{-\frac{1}{2}|K_{\hat{N}_c}|^2(\bar{n}_c+\frac{1}{2})(n-m+p-q)^2 } e^{-\frac{1}{2}|K_{\hat{N}_s}|^2(\bar{n}_s+\frac{1}{2})(m+n-p-q)^2}, 
\end{aligned}
\end{equation}
where the local actions on the optical states in (\ref{Ugeneral}) have been ignored. Further simplifications can also be made if we restrict our attention to short times around the optimal measurement time, $t_q -\Delta t \leq t \leq t_q+\Delta t$. This is consistent with our expectation of a small measurement window, and for $\bar{n}_c=\bar{n}_s=\bar{n}$, an optimal strategy is to choose the modulation frequency $\Omega_n=(1-1/n)(\omega_c+\omega_s)/2$, with $t_q=qn\pi(1/\omega_c+1/\omega_s)$. The difference between $|K_{\hat{N}_s}(t)|^2$ and $|K_{\hat{N}_c}(t)|^2$ is then proportional to $\gamma_g\Delta t$, and so to lowest order in $\gamma_g$, we can assume that 
\begin{equation}
\begin{aligned}\label{eq:MredTheta}
\tilde{M}_{mpnq}\approx & \,a_m a_p^* b_n b_q^* e^{iD(mn-pq)} e^{-\kappa ((m - p)^2 + (n - q)^2) } , 
\end{aligned}
\end{equation}
where $\kappa = \kappa_c^\mathrm{th}= |K_{\hat{N}_c}|^2(\bar{n}+\frac{1}{2})$.

For initial states described by $a_m=b_m=(\delta^0_m+\delta^N_m)/\sqrt{2}$, we can map the partial transpose of the cavity state,
\begin{equation}
\begin{aligned}
\tilde{\rho}^{T_2}_{cav}\left(t\right)&=\sum_{m,n,p,q\in\{0,N\} } \tilde{M}_{mpqn} \ketbra{m}{p}\ketbra{n}{q},
\end{aligned}
\end{equation}
to its $4\times4$ matrix representation $(\tilde{\rho}^{T_2}_{cav}(t))_{(m,n),(p,q)}=\tilde{M}_{mpqn}$, where the indices run over $(j,k) = (0,0),(N,0),(0,N),(N,N)$. The trace norm is then found from the sum of the eigenvalues of $(\tilde{\rho}^{T_2}_{cav}(t))^\dagger (\tilde{\rho}^{T_2}_{cav}(t))$, leading to,
\begin{equation}\label{eq:LNtapp}
	LN(\rho_{cav}(t)) \approx   \left\{
	\begin{array}{ll} 
	    0  & :\, \sinh(\kappa N^2) \ge \sin\left(\frac{D N^2}{2}\right) \\
	    \log\left[e^{-\kappa N^2}\left(\cosh(\kappa N^2) + \sin\left(\frac{D N^2}{2}\right)\right)\right] & :\, \rm{else}
	\end{array}             
	\right. .
\end{equation}
This expression shows that the logarithmic negativity is non-vanishing only in regions where $\sinh(\kappa N^2) < \sin\left(\frac{D N^2}{2}\right)$. These define the available measurement windows in which entanglement can be detected by measurements on the optical fields. In particular, for small arguments of the $\sinh$ and $\sin$ functions, the condition takes the form $\kappa \le D/2$ which coincides with equation (\ref{kappalimit}). A repeat of the analysis leading to (\ref{windowthermalopt}) is straightforward, though in this case $\kappa^{\rm{th}}_{\max} \approx \sinh^{-1}(1)/N^2$ and so from (\ref{kappawindow}),
\begin{equation}
\delta t \approx \frac{\sqrt{2 \sinh^{-1}(1)}}{\omega_m k_0 N \sqrt{\bar{n}+1/2}}.
\end{equation}
Note that as one can expect the optimal measurement time now scales as $t_q \sim 1/N^2$ (see Appendix \ref{App:LinearEntropy}), for both coherent states and Fock state superpositions the measurement window is inversely proportional to the square root of $t$, i.e. its behaviour is relatively insensitive to the initial state.

\begin{figure}[htb]
	\subfloat[]{\includegraphics[width=8cm,angle=0]{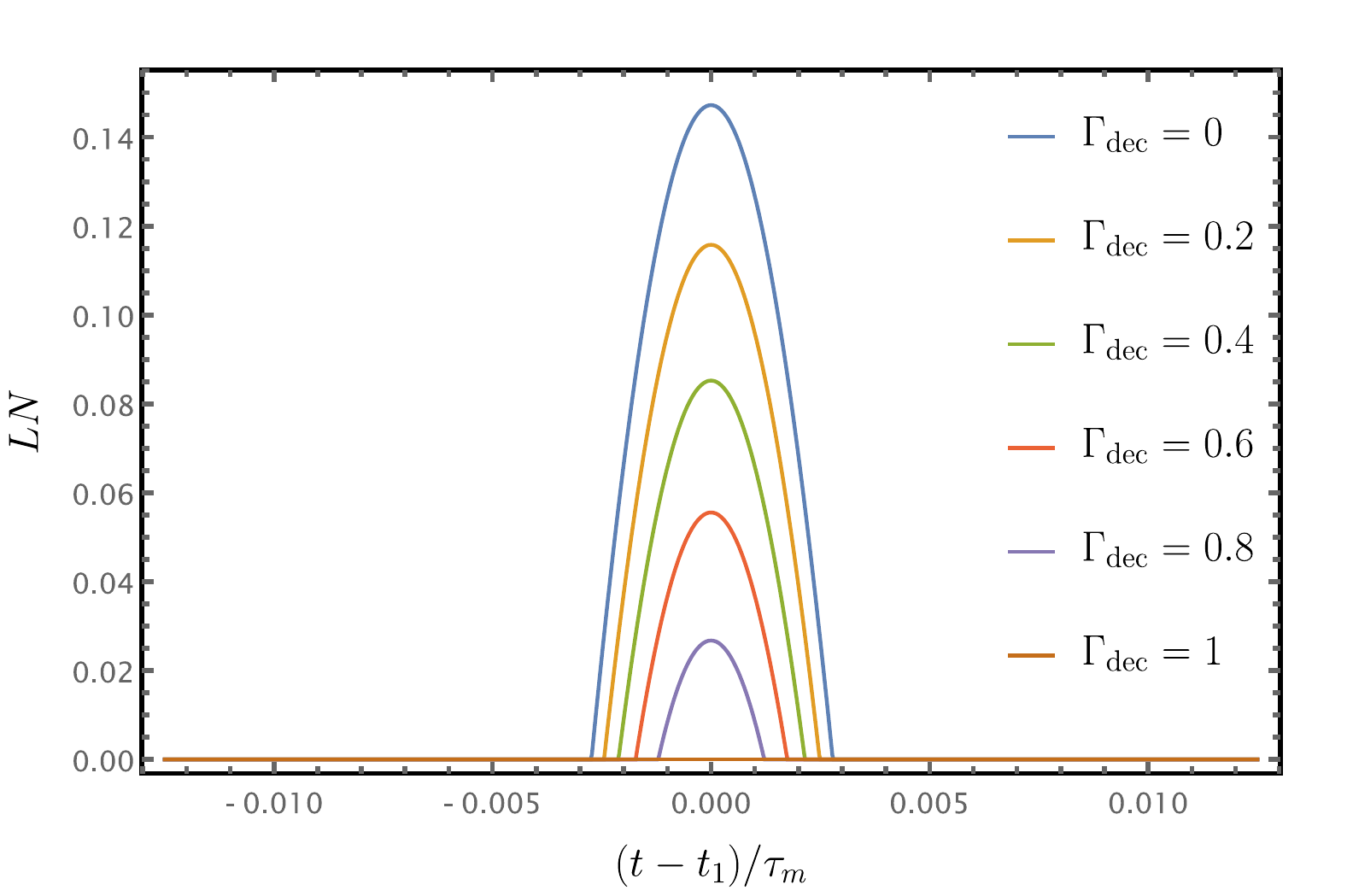}}
		\hspace{0.5cm}\subfloat[]{	\includegraphics[width=8cm,angle=0]{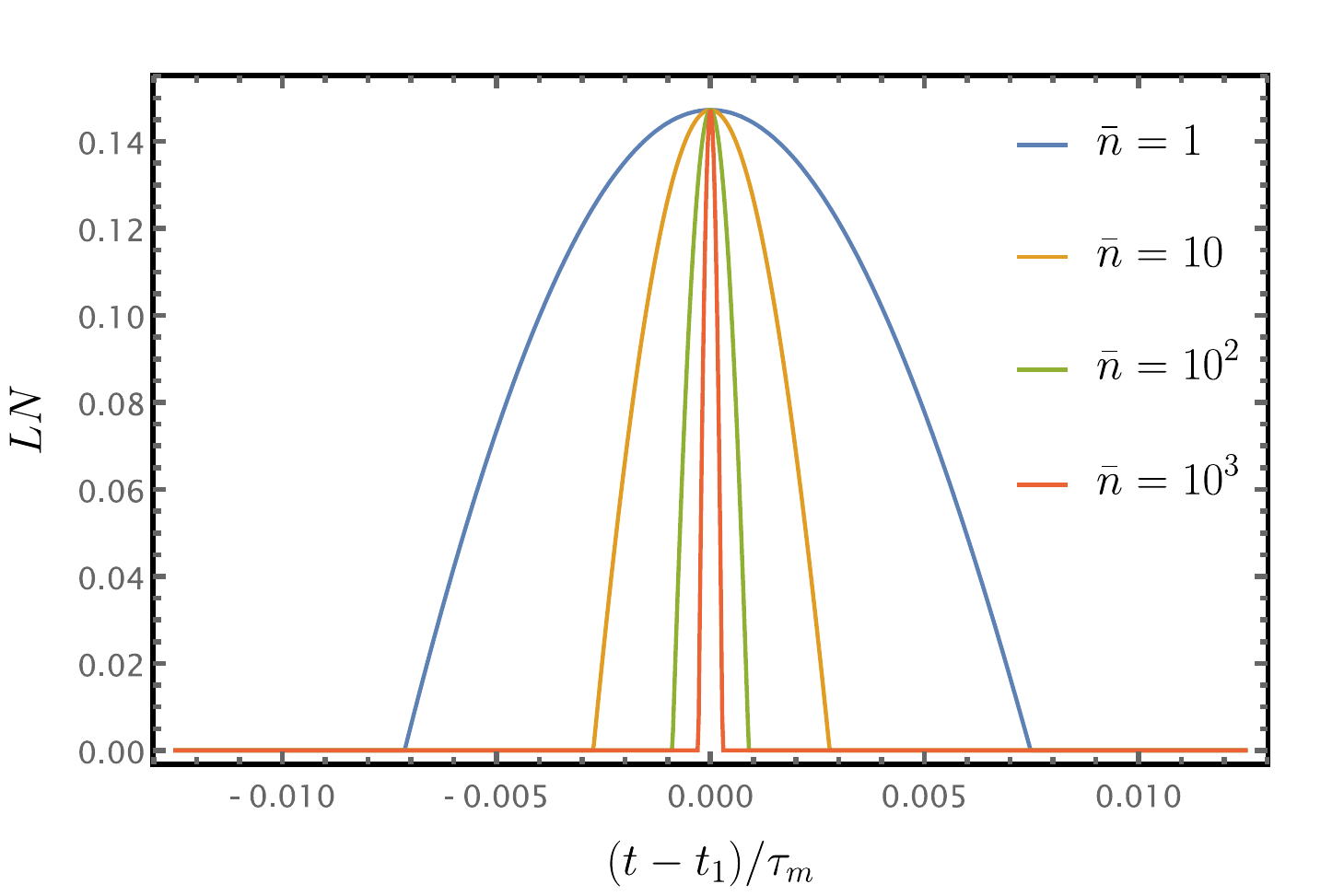}}
	\caption{\label{LNfull} The logarithmic negativity of initial cat states $\ket{\psi_{1,2}}=\frac{1}{\sqrt{2}}(\ket{0}+\ket{10})$ for optomechanical couplings modulated at the fractional frequency  $\Omega_n=(1-1/n)(\omega_c+\omega_s)/2$, $n=1000$ (and phase $\phi_1=\pi/2$) as a function of the time-difference to the  measurement time $t_1=n\pi(1/\omega_c+1/\omega_s)$, in units of $\tau_m=2\pi/\omega_m$. Here we choose the more optimistic parameters, $\omega_m=2 \pi \times 10^2$Hz and $m=2.4\times 10^{-8}$kg, leading to $\gamma_g = 5 \times 10^{-13}$.
	Plot a) shows $LN$ 	for different values of the thermal occupation of the mechanical system $\bar{n}$ and plot b) shows $LN$ for different decoherence rates quantified by the parameter  $\Gamma_\mathrm{dec}$ for $\bar{n}=10$.
}
\end{figure}

The above analysis can be repeated to include decoherence via the substitution $\kappa_\mu \rightarrow \kappa_\mu=\kappa^{\rm{th}}_\mu + \kappa^{\rm{dec}}_\mu$ in (\ref{thermaldensityredcoeff}). Just as with the thermal contributions, for small times around $t_q$ the difference between $\kappa_s^\mathrm{dec}$ and $\kappa_c^\mathrm{dec}$ is proportional to $\gamma_g$, thus \eqref{eq:MredTheta} is again fulfilled in this regime, which immediately recovers equation \eqref{eq:LNtapp} with $\kappa= \kappa_c^\mathrm{th} + \kappa_c^\mathrm{dec}$.

Figure \ref{LNfull} shows plots of the logarithmic negativity close to $t_1$ for the analytic result above. These agree with numerical evaluations to within a relative error of less than $10^{-4}$ everywhere up to regions close to the boundary of the measurement windows (where the numerical errors becomes significant). For the case of decoherence, we find that the measurement window closes as $\Gamma_\mathrm{dec}$ approaches $1$.

%------------------------------------------------------------------------
\section{Displacement and Variance of the Mechanical Modes}\label{sec:displ}
The time dependent position operators for the c.o.m. and stretched modes can be calculated as in Appendix \ref{App:Noise}. We find, 
\begin{equation}
\begin{aligned}\label{App:positions}
\hat{x}_c(t)&=\hat{x}_{\omega_c}(t) + 2 x_{0,c}\Re [ e^{-i\omega_s t} K_{\hat{N}_c}](\hat{N}_1-\hat{N}_2),\\
\hat{x}_s(t)&=\hat{x}_{\omega_s}(t) + 2 x_{0,s}\left(\Re [e^{-i\omega_s t} C_s]  - \Re [ e^{-i\omega_s t} K_{\hat{N}_s}](\hat{N}_1+\hat{N}_2)\right),
\end{aligned}
\end{equation}
where, 
\begin{equation}
\begin{aligned}
\hat{x}_{\omega_\mu}(t)&=x_{0,\mu}\left(e^{i\omega_\mu t}\bd_\mu + e^{-i\omega_\mu t}\hat{b}_\mu\right)\\
&= \hat{x}_\mu(0)\cos \omega_\mu t + \frac{1}{m \omega_m}\hat{p}_\mu(0)\sin \omega_\mu t\,.
\end{aligned}
\end{equation}
The stretched mode gives information of the separation of the two mechanical elements via $\hat{x}_2-\hat{x}_1=\sqrt{2}\hat{x}_s \equiv \delta \hat{x}$. For thermal states, we have that the separation expectation value is simply,
\begin{equation}\label{App:mean_separation}
 d + \avg{\delta \hat{x}} = d +  2 \sqrt{2} x_{0,s}\left(\Re [ e^{-i\omega_s t} C_s] - \Re[ e^{-i\omega_s t} K_{\hat{N}_s}](\avg{\hat{N}_1}+\avg{\hat{N}_2})\right).
\end{equation}
Similarly, the variance in the mode operators is given by, 
\begin{equation}
\begin{aligned}
(\Delta \hat{x}_c)^2 &= (\Delta \hat{x}_{\omega_c})^2 + 4 x_{0,c}^2 \Re[ e^{-i\omega_s t} K_{\hat{N}_c}]^2 \left(\Delta (\hat{N}_1-\hat{N}_2)\right)^2 , \\ (\Delta \hat{x}_s)^2&= (\Delta \hat{x}_{\omega_s})^2 + 4 x_{0,s}^2 \Re [ e^{-i\omega_s t} K_{\hat{N}_s}]^2 \left(\Delta (\hat{N}_1+\hat{N}_2)\right)^2 .
\end{aligned}
\end{equation}
Now, for a thermal state, $(\Delta \hat{x}_{\omega_\mu})^2 = (2\bar{n}_\mu+1)x_{0,\mu}^2$, thus the variance in separation is,
\begin{equation}
(\Delta (d+\delta\hat{x}))^2=2x_{0,s}^2\left[1+2\bar{n}_s + 4\Re[ e^{-i\omega_s t} K_{\hat{N}_s}]^2 \left(\Delta (\hat{N}_1+\hat{N}_2)\right)^2 \right].
\end{equation}

\end{document}

%% file: twocavitysuperpositionalpha.pdf_tex
%% Accompanies image file 'twocavitysuperpositionalpha.pdf' (pdf, eps, ps)
\begingroup%
  \makeatletter%
  \providecommand\color[2][]{%
    \errmessage{(Inkscape) Color is used for the text in Inkscape, but the package 'color.sty' is not loaded}%
    \renewcommand\color[2][]{}%
  }%
  \providecommand\transparent[1]{%
    \errmessage{(Inkscape) Transparency is used (non-zero) for the text in Inkscape, but the package 'transparent.sty' is not loaded}%
    \renewcommand\transparent[1]{}%
  }%
  \providecommand\rotatebox[2]{#2}%
  \newcommand*\fsize{\dimexpr\f@size pt\relax}%
  \newcommand*\lineheight[1]{\fontsize{\fsize}{#1\fsize}\selectfont}%
  \ifx\svgwidth\undefined%
    \setlength{\unitlength}{595.27559055bp}%
    \ifx\svgscale\undefined%
      \relax%
    \else%
      \setlength{\unitlength}{\unitlength * \real{\svgscale}}%
    \fi%
  \else%
    \setlength{\unitlength}{\svgwidth}%
  \fi%
  \global\let\svgwidth\undefined%
  \global\let\svgscale\undefined%
  \makeatother%
  \begin{picture}(1,0.30952381)%
    \lineheight{1}%
    \setlength\tabcolsep{0pt}%
    \put(0,0){\includegraphics[width=\unitlength,page=1]{twocavitysuperpositionalpha.pdf}}%
    \put(0.38208248,0.02077128){\color[rgb]{0,0,0}\makebox(0,0)[lt]{\lineheight{1.25}\smash{\begin{tabular}[t]{l}\end{tabular}}}}%
    \put(0.5205513,0.02013492){\color[rgb]{0,0,0}\makebox(0,0)[lt]{\lineheight{1.25}\smash{\begin{tabular}[t]{l}\end{tabular}}}}%
    \put(0.21553643,0.171176){\color[rgb]{0,0,0}\makebox(0,0)[lt]{\lineheight{1.25}\smash{\begin{tabular}[t]{l}$\omega_0 $\end{tabular}}}}%
    \put(0.0252055,0.07076939){\color[rgb]{0,0,0}\makebox(0,0)[lt]{\lineheight{1.25}\smash{\begin{tabular}[t]{l}\end{tabular}}}}%
    \put(0,0){\includegraphics[width=\unitlength,page=2]{twocavitysuperpositionalpha.pdf}}%
    \put(0.46369854,0.28455375){\color[rgb]{0,0,0}\makebox(0,0)[lt]{\lineheight{1.25}\smash{\begin{tabular}[t]{l}$d$\end{tabular}}}}%
    \put(0.38823042,0.20959436){\color[rgb]{0,0,0}\makebox(0,0)[lt]{\lineheight{1.25}\smash{\begin{tabular}[t]{l}$x_1$\end{tabular}}}}%
    \put(0,0){\includegraphics[width=\unitlength,page=3]{twocavitysuperpositionalpha.pdf}}%
    \put(0.23625906,0.12920227){\color[rgb]{0,0,0}\makebox(0,0)[lt]{\lineheight{1.25}\smash{\begin{tabular}[t]{l}$L$\end{tabular}}}}%
    \put(0.57240608,0.23500337){\color[rgb]{0,0,0}\makebox(0,0)[lt]{\lineheight{1.25}\smash{\begin{tabular}[t]{l}$0$\end{tabular}}}}%
    \put(0,0){\includegraphics[width=\unitlength,page=4]{twocavitysuperpositionalpha.pdf}}%
    \put(0.35513989,0.23611225){\color[rgb]{0,0,0}\makebox(0,0)[lt]{\lineheight{1.25}\smash{\begin{tabular}[t]{l}$0$\end{tabular}}}}%
    \put(0.60757657,0.20957177){\color[rgb]{0,0,0}\makebox(0,0)[lt]{\lineheight{1.25}\smash{\begin{tabular}[t]{l}$x_2$\end{tabular}}}}%
    \put(0,0){\includegraphics[width=\unitlength,page=5]{twocavitysuperpositionalpha.pdf}}%
  \end{picture}%
\endgroup%